\documentclass[paper]{JHEP}
\usepackage{epsfig}
\title{\boldmath  
Model-Independent Analysis of $B\to\pi K$ Decays and Bounds on the 
Weak Phase $\gamma$
\unboldmath}

\author{
Matthias Neubert\thanks{Address after 1 January 1999: 
Theory Group, Stanford Linear Accelerator Center, Stanford 
University, Stanford, California 94309, U.S.A.}\\
Theory Division, CERN, CH-1211 Geneva 23, Switzerland\\
E-mail: \email{neubert@slac.stanford.edu}}

\abstract{A general parametrization of the amplitudes for the rare 
two-body decays $B\to\pi K$ is introduced, which makes maximal use 
of theoretical constraints arising from flavour symmetries of the 
strong interactions and the structure of the low-energy effective 
weak Hamiltonian. With the help of this parametrization, 
a model-independent analysis of the branching ratios and direct CP 
asymmetries in the various $B\to\pi K$ decay modes is performed, 
and the impact of hadronic uncertainties on bounds on the weak 
phase $\gamma=\mbox{arg}(V_{ub}^*)$ is investigated.}


\preprint{CERN-TH/98-384\\
hep-ph/9812396}

\begin{document}

\section{Introduction}

The CLEO Collaboration has recently reported the observation of
some rare two-body decays of the type $B\to\pi K$, as well as 
interesting upper bounds for the decays $B\to\pi\pi$ and $B\to 
K\bar K$ \cite{CLEO}. In particular, they find the CP-averaged
branching ratios
\begin{eqnarray}
   \frac 12 \Big[ \mbox{Br}(B^0\to\pi^- K^+) 
   + \mbox{Br}(\bar B^0\to\pi^+ K^-) \Big]
   &=& (1.4\pm 0.3\pm 0.1)\times 10^{-5} \,, \nonumber\\
   \frac 12 \Big[ \mbox{Br}(B^+\to\pi^+ K^0)
   + \mbox{Br}(B^-\to\pi^-\bar K^0) \Big]
   &=& (1.4\pm 0.5\pm 0.2)\times 10^{-5} \,, \nonumber\\
   \frac 12 \Big[ \mbox{Br}(B^+\to\pi^0 K^+)
   + \mbox{Br}(B^-\to\pi^0 K^-) \Big]
   &=& (1.5\pm 0.4\pm 0.3)\times 10^{-5} \,.
\label{CLEOvals}
\end{eqnarray}
This observation caused a lot of excitement, because these 
decays offer interesting insights into the relative strength of 
various contributions to the decay amplitudes, whose interference 
can lead to CP asymmetries in the decay rates. It indeed appears 
that there may be potentially large interference effects, 
depending on the magnitude of some strong interaction phases 
(see, e.g., \cite{GrRo}). Thus, although at present only 
measurements of CP-averaged branching ratios have been reported, 
the prospects are good for observing direct CP violation in some 
of the $B\to\pi K$ or $B\to K\bar K$ decay modes in the near 
future. 

It is fascinating that some information on CP-violating parameters 
can be extracted even without observing a single CP asymmetry, from 
measurements of CP-averaged branching ratios alone. This information
concerns the angle $\gamma$ of the so-called unitarity triangle, 
defined as $\gamma=\mbox{arg}[(V_{ub}^* V_{ud})/(V_{cb}^* V_{cd})]$.
With the standard phase conventions for the 
Cabibbo--Kobayashi--Maskawa (CKM) matrix, $\gamma=
\mbox{arg}(V_{ub}^*)$ to excellent accuracy. There have been
proposals for deriving bounds on $\gamma$ from measurements of the
ratios
\begin{eqnarray}
   R &=& \frac{\tau(B^+)}{\tau(B^0)}\,  
    \frac{\mbox{Br}(B^0\to\pi^- K^+)+\mbox{Br}(\bar B^0\to\pi^+ K^-)}
         {\mbox{Br}(B^+\to\pi^+ K^0)+\mbox{Br}(B^-\to\pi^-\bar K^0)}
    \,, \nonumber\\
   R_* &=&
    \frac{\mbox{Br}(B^+\to\pi^+ K^0)+\mbox{Br}(B^-\to\pi^-\bar K^0)}
         {2[\mbox{Br}(B^+\to\pi^0 K^+)+\mbox{Br}(B^-\to\pi^0 K^-)]} \,,
\label{Rdef}
\end{eqnarray}
whose current experimental values are $R=1.07\pm 0.45$ (we use 
$\tau(B^+)/\tau(B^0)=1.07\pm 0.03$) and $R_*=0.47\pm 0.24$. The 
Fleischer--Mannel bound $R\ge\sin^2\!\gamma$ \cite{FM} excludes 
values around $|\gamma|=90^\circ$ provided that $R<1$. However, 
this bound is subject to theoretical uncertainties arising from 
electroweak penguin contributions and strong rescattering effects, 
which are difficult to quantify \cite{BFM98}--\cite{Robert}. The 
bound
\begin{equation}
   1-\sqrt{R_*} \le \bar\varepsilon_{3/2}\,
   |\delta_{\rm EW} - \cos\gamma| + O(\bar\varepsilon_{3/2}^2)
\label{ourb}
\end{equation}
derived by Rosner and the present author \cite{us}, where 
$\delta_{\rm EW}=0.64\pm 0.15$ accounts for electroweak penguin
contributions, is less affected by such uncertainties; however, it 
relies on an expansion in the small parameter
\begin{equation}
   \bar\varepsilon_{3/2} = \sqrt 2\,R_{\rm SU(3)} \tan\theta_C \left[
   \frac{\mbox{Br}(B^+\to\pi^+\pi^0) + \mbox{Br}(B^-\to\pi^-\pi^0)}
        {\mbox{Br}(B^+\to\pi^+ K^0) + \mbox{Br}(B^-\to\pi^-\bar K^0)}
   \right]^{1/2} \,, 
\label{epsexp}
\end{equation}
whose value has been estimated to be $\bar\varepsilon_{3/2}
=0.24\pm 0.06$. Here $\theta_C$ is the Cabibbo angle, and the factor 
$R_{\rm SU(3)}\simeq f_K/f_\pi$ accounts for SU(3)-breaking 
corrections. Assuming the smallness of certain rescattering effects,
higher-order terms in the expansion in $\bar\varepsilon_{3/2}$ can be 
shown to strengthen the bound (\ref{ourb}) provided that the value of 
$R_*$ is not much larger than indicated by current data, i.e., if 
$R_*<(1-\bar\varepsilon_{3/2}/\sqrt 2)^2\approx 0.7$ \cite{us}. 

Our main goal in the present work is to address the question to what 
extent these bounds can be affected by hadronic uncertainties such as 
final-state rescattering effects, and whether the theoretical 
assumptions underlying them are justified. To this end, we  
perform a general analysis of the various $B\to\pi K$ 
decay modes, pointing out where theoretical information from isopsin
and SU(3) flavour symmetries can be used to eliminate hadronic 
uncertainties. Our approach will be to vary parameters not 
constrained by theory (strong-interaction phases, in particular) 
within conservative ranges so as to obtain a model-independent 
description of the decay amplitudes. An analysis pursuing a similar 
goal has recently been presented by Buras and Fleischer \cite{BFnew}. 
Where appropriate, we will point out the relations of our work with 
theirs and provide a translation of notations. We stress, however, 
that although we take a similar starting point, some of our 
conclusions will be rather different from the ones reached in their 
work. 

In Section~\ref{sec:2}, we present a general parametrization of the 
various isospin amplitudes relevant to $B\to\pi K$ decays and discuss
theoretical constraints resulting from flavour symmetries of the 
strong interactions and the structure of the low-energy 
effective weak Hamiltonian. We summarize model-independent results
derived recently for the electroweak penguin contributions to the
isovector part of the effective Hamiltonian \cite{Ne97,Robert} and 
point out constraints on certain rescattering contributions
resulting from $B\to K\bar K$ decays \cite{Fa97,Robert,GRrescat,He}.
The main results of this analysis are presented in 
Section~\ref{subsec:numerics}, which contains numerical predictions
for the various parameters entering our parametrization of the 
decay amplitudes. The remainder of the paper deals with 
phenomenological applications of these results. In 
Section~\ref{sec:3}, we discuss corrections to the Fleischer--Mannel 
bound resulting from final-state rescattering and electroweak 
penguin contributions. In Section~\ref{sec:4}, we show how to
include rescattering effects to the bound (\ref{ourb}) at 
higher orders in the expansion in $\bar\varepsilon_{3/2}$. Detailed 
predictions for the direct CP asymmetries in the various $B\to\pi K$
decay modes are presented in Section~\ref{sec:5}, where we also
present a prediction for the CP-averaged $B^0\to\pi^0 K^0$ branching 
ratio, for which at present only an upper limit exists. In 
Section~\ref{sec:6}, we discuss how the weak phase $\gamma$, 
along with a strong-interaction phase difference $\phi$, can be 
determined from measurements of the ratio $R_*$ and 
of the direct CP asymmetries in the decays $B^\pm\to\pi^0 K^\pm$ and 
$B^\pm\to\pi^\pm K^0$ (here $K^0$ means $K^0$ or $\bar K^0$, as 
appropriate). This generalizes a method proposed in
\cite{us2} to include rescattering corrections to the 
$B^\pm\to\pi^\pm K^0$ decay amplitudes. Section~\ref{sec:7} contains
a summary of our result and the conclusions.

\section{Isospin decomposition}
\label{sec:2}

\subsection{Preliminaries}

The effective weak Hamiltonian relevant to the decays $B\to\pi K$ is 
\cite{Heff}
\begin{equation}
   {\mathcal H} = \frac{G_F}{\sqrt 2}\,\bigg\{ \sum_{i=1,2} C_i \Big(
   \lambda_u\,Q_i^u + \lambda_c\,Q_i^c \Big) - \lambda_t
   \sum_{i=3}^{10} C_i\,Q_i  \bigg\} + \mbox{h.c.} \,,
\end{equation}
where $\lambda_q=V_{qb}^* V_{qs}$ are products of CKM matrix elements,
$C_i$ are Wilson coefficients, and $Q_i$ are local four-quark
operators. Relevant to our discussion are the isospin quantum
numbers of these operators. The current--current operators
$Q_{1,2}^u\sim\bar b s\bar u u$ have components with $\Delta I=0$ and
$\Delta I=1$; the current--current operators  $Q_{1,2}^c\sim\bar b
s\bar c c$ and the QCD penguin operators $Q_{3,\dots,6} \sim \bar b
s\sum\bar q q$ have $\Delta I=0$; the electroweak penguin operators
$Q_{7,\dots,10}\sim\bar b s \sum e_q\bar q q$, where $e_q$ are the
electric charges of the quarks, have $\Delta I=0$ and $\Delta
I=1$. Since the initial $B$ meson has $I=\frac 12$ and the final
states $(\pi K)$ can be decomposed into components with $I=\frac 12$
and $I=\frac 32$, the physical $B\to\pi K$ decay amplitudes can be
described in terms of three isospin amplitudes. They are called
$B_{1/2}$, $A_{1/2}$, and $A_{3/2}$ referring, respectively, to
$\Delta I=0$ with $I_{\pi K}=\frac 12$, $\Delta I=1$ with $I_{\pi
K}=\frac 12$, and  $\Delta I=1$ with $I_{\pi K}=\frac 32$
\cite{Ne97,Gron,NQ}. The resulting expressions for the decay 
amplitudes are
\begin{eqnarray}
   {\mathcal A}(B^+\to\pi^+ K^0)
   &=& B_{1/2} + A_{1/2} + A_{3/2} \,, \nonumber\\
   - \sqrt 2\,{\mathcal A}(B^+\to\pi^0 K^+)
   &=& B_{1/2} + A_{1/2} - 2 A_{3/2} \,, \nonumber\\
   - {\mathcal A}(B^0\to\pi^- K^+)
   &=& B_{1/2} - A_{1/2} - A_{3/2} \,, \nonumber\\
   \sqrt 2\,{\mathcal A}(B^0\to\pi^0 K^0)
   &=& B_{1/2} - A_{1/2} + 2 A_{3/2} \,.
\label{isodec}
\end{eqnarray}
From the isospin decomposition of the effective Hamiltonian it is 
obvious which operator matrix elements and weak phases enter the 
various isospin amplitudes. Experimental data as well as theoretical 
expectations indicate that the amplitude $B_{1/2}$, which includes 
the contributions of the QCD penguin operators, is significantly 
larger than the amplitudes $A_{1/2}$ and $A_{3/2}$ \cite{GrRo,Ne97}. 
Yet, the fact that $A_{1/2}$ and $A_{3/2}$ are different from zero 
is responsible for the deviations of the ratios $R$ and $R_*$ in 
(\ref{Rdef}) from 1. 

Because of the unitarity relation $\lambda_u+\lambda_c+\lambda_t=0$ 
there are two independent CKM parameters entering the decay
amplitudes, which we choose to be\footnote{Taking $\lambda_c$ to be
real is an excellent approximation.} 
$-\lambda_c=e^{i\pi}|\lambda_c|$ and 
$\lambda_u=e^{i\gamma}|\lambda_u|$. Each of the three isospin 
amplitudes receives contributions proportional to both weak phases.
In total, there are thus five independent strong-interaction phase
differences (an overall phase is irrelevant) and six independent
real amplitudes, leaving as many as eleven hadronic parameters. 
Even perfect measurements of the eight branching ratios for the 
various $B\to\pi K$ decay modes and their CP conjugates would not
suffice to determine these parameters. Facing this problem, previous
authors have often relied on some theoretical prejudice about the
relative importance of various parameters. For instance, in 
the invariant SU(3)-amplitude approach based on flavour-flow 
topologies \cite{Chau,ampl}, the isospin amplitudes are expressed 
as linear combinations of a QCD penguin amplitude $P$, a tree 
amplitude $T$, a colour-suppressed tree amplitude $C$, an 
annihilation amplitide $A$, an electroweak penguin amplitude 
$P_{\rm EW}$, and a colour-suppressed electroweak penguin amplitude 
$P_{\rm EW}^C$, which are expected to obey the following 
hierarchy: $|P|\gg |T|\sim |P_{\rm EW}|\gg |C|\sim |P_{\rm EW}^C| 
>|A|$. These naive expectations could be upset, however, if 
strong final-state rescattering effects would turn out to be 
important \cite{Ge97}--\cite{At97}, a possibility which at present
is still under debate. Whereas the colour-transparency argument 
\cite{Bj} suggests that final-state interactions are small in $B$ 
decays into a pair of light mesons, the opposite behaviour is 
exhibited in a model based on Regge phenomenology 
\cite{Dono}. For comparison, we note that in the decays 
$B\to D^{(*)} h$, with $h=\pi$ or $\rho$, the final-state phase 
differences between the $I=\frac 12$ and $I=\frac 32$ isospin 
amplitudes are found to be smaller than $30^\circ$--$50^\circ$ 
\cite{Stech}. 

Here we follow a different strategy, making maximal use of 
theoretical constraints derived using flavour symmetries and the 
knowledge of the effective weak Hamiltonian in the Standard Model. 
These constraints help simplifying the isospin amplitude 
$A_{3/2}$, for which the two contributions with different weak 
phases turn out to have the same strong-interaction phase (to an 
excellent approximation) and magnitudes that can be determined 
without encountering large hadronic uncertainties \cite{us}. 
Theoretical uncertainties enter only at the level of 
SU(3)-breaking corrections, which can be accounted for using the 
generalized factorization approximation \cite{Stech}. Effectively, 
these simplifications remove three parameters (one phase and two 
magnitudes) from the list of unknown hadronic quantities. 
There is at present no other clean theoretical 
information about the remaining parameters, although some 
constraints can be derived using measurements of the branching 
ratios for the decays $B^\pm\to K^\pm\bar K^0$ and invoking SU(3) 
symmetry \cite{Fa97,Robert,GRrescat,He}. Nevertheless, interesting 
insights can be gained by fully exploiting the available 
information on $A_{3/2}$. 

Before discussing this in more detail, it is instructive to 
introduce certain linear combinations of the isospin amplitudes, 
which we define as
\begin{eqnarray}
   B_{1/2} + A_{1/2} + A_{3/2} &=& P + A - \frac 13 P_{\rm EW}^C
    \,, \nonumber\\
   -3 A_{3/2} &=& T + C + P_{\rm EW} + P_{\rm EW}^C \,, \nonumber\\
   -2(A_{1/2} + A_{3/2}) &=& T - A + P_{\rm EW}^C \,.
\label{combo}
\end{eqnarray}
In the latter two relations, the amplitudes $T$, $C$ and $A$
carry the weak phase $e^{i\gamma}$, whereas the electroweak penguin
amplitudes $P_{\rm EW}$ and $P_{\rm EW}^C$ carry the weak 
phase\footnote{Because of their smallness, it is a safe approximation
to set $\lambda_t=-\lambda_c$ for the electroweak penguin 
contributions, and to neglect electroweak penguin contractions 
in the matrix elements of the four-quark operators $Q_i^u$ and 
$Q_i^c$.} 
$e^{i\pi}$. Decomposing the QCD penguin amplitude as $P=\sum_q 
\lambda_q P_q$, and similarly writing $A=\lambda_u A_u$ and 
$P_{\rm EW}^C=\lambda_t P_{{\rm EW},t}^C$, we rewrite the first 
relation in the form
\begin{eqnarray}
   B_{1/2} + A_{1/2} + A_{3/2}
   &=& - \lambda_c (P_t - P_c - \textstyle\frac 13 P_{{\rm EW},t}^C)
    + \lambda_u (A_u - P_t + P_u) \nonumber\\
   &\equiv& |P|\,e^{i\phi_P} \Big( e^{i\pi}
    + \varepsilon_a\,e^{i\gamma} e^{i\eta} \Big) \,.
\label{ampl1}
\end{eqnarray}
By definition, the term $|P|\,e^{i\phi_P} e^{i\pi}$ contains all 
contributions to the $B^+\to\pi^+ K^0$ decay amplitude not 
proportional to the weak phase $e^{i\gamma}$. We will return to a
discussion of the remaining terms below. It is convenient to adopt
a parametrization of the other two amplitude combinations in 
(\ref{combo}) in units of $|P|$, so that this parameter cancels in 
predictions for ratios of branching ratios. We define
\begin{eqnarray}
   \frac{-3 A_{3/2}}{|P|} &=& \varepsilon_{3/2}\,e^{i\phi_{3/2}}
    (e^{i\gamma} - q\,e^{i\omega}) \,, \nonumber\\
   \frac{-2(A_{1/2}+A_{3/2})}{|P|} &=& \varepsilon_T\,
    e^{i\phi_T} (e^{i\gamma} - q_C\,e^{i\omega_C}) \,, 
\label{ampl2}
\end{eqnarray}
where the terms with $q$ and $q_C$ arise from electroweak penguin
contributions. In the above relations, the parameters $\eta$, 
$\phi_{3/2}$, $\phi_T$, $\omega$, and $\omega_C$ are 
strong-interaction phases. For the benefit of the reader, it may be 
convenient
to relate our definitions in (\ref{ampl1}) and (\ref{ampl2}) with 
those adopted by Buras and Fleischer \cite{BFnew}. The 
identificantions are: $|P|\,e^{i\phi_P}\leftrightarrow
\lambda_c|P_{tc}|\,e^{i\delta_{tc}}$, 
$\varepsilon_a\,e^{i\eta}\leftrightarrow -\rho\,e^{i\theta}$, 
$\phi_{3/2}\leftrightarrow\delta_{T+C}$, and
$\phi_T\leftrightarrow\delta_T$. The notations for the electroweak 
penguin contributions conincide. Moreover, if we define
\begin{equation}
   \bar\varepsilon_{3/2}\equiv \frac{\varepsilon_{3/2}}
    {\sqrt{1-2\varepsilon_a\cos\eta\cos\gamma+\varepsilon_a^2}} \,,
\label{bar32}
\end{equation}
then $\bar\varepsilon_{3/2}\leftrightarrow r_{\rm c}$ and
$\varepsilon_T/\varepsilon_{3/2}\leftrightarrow r/r_{\rm c}$. With
this definition, the parameter $\bar\varepsilon_{3/2}$ is precisely
the quantity that can be determined experimentally using the relation 
(\ref{epsexp}).

\subsection{Isovector part of the effective weak Hamiltonian}

The two amplitude combinations in (\ref{ampl2}) involve isospin 
amplitudes defined in terms of the strong-interaction matrix elements
of the $\Delta I=1$ part of the effective weak 
Hamiltonian.\footnote{This statement implies that QED corrections to
the matrix elements are neglected, which is an excellent 
approximation.}
This part contains current--current as well as electroweak penguin
operators. A trivial but relevant observation is that the electroweak 
penguin operators $Q_9$ and $Q_{10}$, whose Wilson coefficients are 
enhanced by the large mass of the top quark, are Fierz-equivalent to 
the current--current operators $Q_1$ and $Q_2$ 
\cite{Ne97,Robert,Fl96}. As a result, the $\Delta I=1$ part of the 
effective weak Hamiltonian for $B\to\pi K$ decays can be written as
\begin{equation}
   {\mathcal H}_{\Delta I=1} = \frac{G_F}{\sqrt 2} \left\{ \left(
   \lambda_u C_1 - \frac 32\lambda_t C_9 \right) \bar Q_1 + \left(
   \lambda_u C_2 - \frac 32\lambda_t C_{10} \right) \bar Q_2 + \dots
   \right\} + \mbox{h.c.} \,,
\label{newH}
\end{equation}
where $\bar Q_i=\frac 12(Q_i^u-Q_i^d)$ are isovector combinations
of four-quark operators. The dots represent the contributions from 
the electroweak penguin operators $Q_7$ and $Q_8$, which have a 
different Dirac structure. In the Standard Model, the Wilson 
coefficients of these operators are so small that their contributions 
can be safely neglected. It is important in this context that for 
heavy mesons the matrix elements of four-quark operators with Dirac 
structure $(V-A)\otimes (V+A)$ are not enhanced with respect to those 
of operators with the usual $(V-A)\otimes(V-A)$ structure. 
To an excellent approximation, the net effect of electroweak penguin
contributions to the $\Delta I=1$ isospin amplitudes in $B\to\pi K$
decays thus consists of the replacements of the Wilson coefficients
$C_1$ and $C_2$ of the current--current operators with the
combinations shown in (\ref{newH}). Introducing the linear 
combinations $C_\pm=(C_2\pm C_1)$ and $\bar Q_\pm=\frac 12
(\bar Q_2\pm\bar Q_1)$, which have the advantage of being 
renormalized multiplicatively, we obtain
\begin{equation}
   {\mathcal H}_{\Delta I=1}\simeq \frac{G_F}{\sqrt 2}
   |V_{ub}^* V_{us}| \left\{ C_+ (e^{i\gamma} - \delta_+)\,\bar Q_+
   + C_- (e^{i\gamma} - \delta_-)\,\bar Q_- \right\} + \mbox{h.c.} \,,
\label{Qpl}
\end{equation}
where
\begin{equation}
   \delta_\pm = - \frac{3\cot\theta_C}{2\,|V_{ub}/V_{cb}|}\,
   \frac{C_{10}\pm C_9}{C_2\pm C_1} \,.
\label{delpm}
\end{equation}
We have used $\lambda_u/\lambda_t\simeq -\lambda_u/\lambda_c
\simeq -\tan\theta_C\,|V_{ub}/V_{cb}|\,e^{i\gamma}$, with the 
ratio $|V_{ub}/V_{cb}|=0.089\pm 0.015$ determined from 
semileptonic $B$ decays \cite{Rosnet}. 

From the fact that the products $C_\pm\,\bar Q_\pm$ are 
renormalization-group invariant, it follows that the quantities 
$\delta_\pm$ themselves must be scheme- and scale-independent (in a 
certain approximation). Indeed, the ratios of Wilson coefficients 
entering in (\ref{delpm}) are, to a good approximation, independent 
of the choice of the renormalization scale. Taking the values 
$C_1=-0.308$, $C_2=1.144$, $C_9=-1.280\alpha$ and 
$C_{10}=0.328\alpha$, which correspond to the leading-order 
coefficients at the scale $\mu=m_b$ \cite{Heff}, we find 
$(C_{10}+C_9)/(C_2+C_1)\approx -1.14\alpha$ 
and $(C_{10}-C_9)/(C_2-C_1)\approx 1.11\alpha$, implying that
$\delta_-\approx -\delta_+$ to a good approximation. 
The statement of the approximate renormalization-group invariance 
of the ratios $\delta_\pm$ can be made more precise by noting that 
the large values of the Wilson coefficients $C_9$ and $C_{10}$ at 
the scale $\mu=m_b$ predominantly result from large matching 
contributions to the coefficient $C_9(m_W)$ arising from box 
and $Z$-penguin diagrams, whereas the $O(\alpha)$ contributions to 
the anomalous dimension matrix governing the mixing of the local 
operators $Q_i$ lead to very 
small effects. If these are neglected, then to next-to-leading 
order in the QCD evolution the coefficients $(C_{10}\pm C_9)$ are 
renormalized multiplicatively and in precisely the same way as the 
coefficients $(C_2\pm C_1)$. We have derived this result using the
explicit expressions for the anomalous dimension matrices compiled
in \cite{Heff}.\footnote{The equivalence of the anomalous dimensions 
at next-to-leading order is nontrivial because the operators $Q_9$ 
and $Q_{10}$ are related to $Q_1$ and $Q_2$ by Fierz identities, 
which are valid only in four dimensions. The corresponding two-loop 
anomalous dimensions are identical in the naive dimensional 
regularization scheme with anticommuting $\gamma_5$.}
Hence, in this approximation the ratios of coefficients entering the 
quantities $\delta_\pm$ are renormalization-scale independent and 
can be evaluated at the scale $m_W$, so that 
\begin{equation}
   \frac{C_{10}\pm C_9}{C_2\pm C_1} \simeq \pm C_9(m_W)
   = \mp \frac{\alpha}{12\pi}\,\frac{x_t}{\sin^2\!\theta_W}
   \left( 1 + \frac{3\ln x_t}{x_t-1} \right) + \dots 
   \approx \mp 1.18\alpha \,,
\label{niceeq}
\end{equation}
where $\theta_W$ is the Weinberg angle, and $x_t=(m_t/m_W)^2$. This 
result agrees with an equivalent expression derived by Fleischer 
\cite{Fl96}. The dots in (\ref{niceeq}) represent 
renormalization-scheme dependent terms, which are not enhanced 
by the factor $1/\sin^2\!\theta_W$. These terms are numerically very 
small and of the same order as the coefficients $C_7$ and $C_8$, 
whose values have been neglected in our derivation. 
The leading terms given above are precisely the ones that must be
kept to get a consistent, renormalization-group invariant result. 
We thus obtain
\begin{equation}
   \delta_+ = - \delta_- = \frac{\alpha}{8\pi}\,
   \frac{\cot\theta_C}{|V_{ub}/V_{cb}|}\,\frac{x_t}{\sin^2\!\theta_W}
   \left( 1 + \frac{3\ln x_t}{x_t-1} \right)
   = 0.68\pm 0.11 \,,
\end{equation}
where we have taken $\alpha=1/129$ for the electromagnetic coupling 
renormalized at the scale $m_b$, and 
$m_t=\overline{m}_t(m_t)=170$\,GeV for the running top-quark mass in 
the $\overline{{\rm MS}}$ renormalization scheme. Assuming that there
are no large $O(\alpha_s)$ corrections with this choice, the main 
uncertainty in the estimate of $\delta_+$ in the Standard Model 
results from the present error on $|V_{ub}|$, which is likely to be 
reduced in the near future. 

We stress that the sensitivity of the $B\to\pi K$ decay amplitudes 
to the value of $\delta_+$ provides a window to New Physics, which 
could alter the value of this parameter significantly. A generic 
example are extensions of the Standard Model with new charged 
Higgs bosons such as supersymmetry, for which there are additional
matching contributions to $C_9(m_W)$. We will come back to this 
point in Section~\ref{sec:4}.

\subsection{Structure of the isospin amplitude \boldmath$A_{3/2}$
\unboldmath}
\label{subsec:A32}

$U$-spin invariance of the strong interactions, which is a subgroup 
of flavour SU(3) symmetry corresponding to transformations exchanging 
$d$ and $s$ quarks, implies that the isospin amplitude $A_{3/2}$ 
receives a contribution only from the operator $\bar Q_+$ in 
(\ref{Qpl}), but not from $\bar Q_-$ \cite{us}. In order to 
investigate the corrections to this limit, we parametrize the matrix 
elements of the local operators $C_\pm \bar Q_\pm$ between a $B$ 
meson and the $(\pi K)$ isospin state with $I=\frac 32$ by hadronic 
parameters $K_{3/2}^\pm\,e^{i\phi_{3/2}^\pm}$, so that
\begin{eqnarray}
   -3 A_{3/2} &=& K_{3/2}^+\,e^{i\phi_{3/2}^+} (e^{i\gamma}-\delta_+)
    + K_{3/2}^-\,e^{i\phi_{3/2}^-} (e^{i\gamma}+\delta_+) \nonumber\\
   &\equiv& \Big( K_{3/2}^+\,e^{i\phi_{3/2}^+} + K_{3/2}^-\, 
    e^{i\phi_{3/2}^-} \Big) (e^{i\gamma} - q\,e^{i\omega}) \,.
\label{Kdef}
\end{eqnarray}
In the SU(3) limit $K_{3/2}^-=0$, and hence SU(3)-breaking corrections 
can be parametrized by the quantity
\begin{equation}
   \kappa\,e^{i\Delta\varphi_{3/2}}
   \equiv \frac{2 K_{3/2}^-\,e^{i\phi_{3/2}^-}}
    {K_{3/2}^+\,e^{i\phi_{3/2}^+} + K_{3/2}^-\,e^{i\phi_{3/2}^-}}
   = 2 \left[ \frac{K_{3/2}^+}{K_{3/2}^-}\,
   e^{i(\phi_{3/2}^+ -\phi_{3/2}^-)} + 1 \right]^{-1} \,,
\label{SU3br}
\end{equation}
in terms of which 
\begin{equation}
   q\,e^{i\omega} = \left( 1 - \kappa\,e^{i\Delta\varphi_{3/2}}
   \right) \delta_+ \,.
\label{dEW}\end{equation}
This relation generalizes an approximate result derived in \cite{us}. 

The magnitude of the SU(3)-breaking effects can be estimated by using 
the generalized factorization hypothesis to calculate the matrix 
elements of the current--current operators \cite{Stech}. This gives 
\begin{equation}
   \kappa\simeq 2 \left[ \frac{a_1+a_2}{a_1-a_2}\,
   \frac{A_K + A_\pi}{A_K - A_\pi} + 1 \right]^{-1}
   = (6\pm 6)\% \,,\qquad \Delta\varphi_{3/2}\simeq 0 \,,
\label{est}
\end{equation}
where $A_K=f_K (m_B^2-m_\pi^2) F_0^{B\to\pi}(m_K^2)$ and $A_\pi=f_\pi
(m_B^2-m_K^2) F_0^{B\to K}(m_\pi^2)$ are combinations of hadronic
matrix elements, and $a_1$ and $a_2$ are phenomenological parameters
defined such that they contain the leading 
corrections to naive factorization. For a numerical estimate we take 
$a_2/a_1=0.21\pm 0.05$ as determined from a global analysis of 
nonleptonic two-body decays of $B$ mesons \cite{Stech}, and 
$A_\pi/A_K=0.9\pm 0.1$, which is consistent with form factor models 
(see, e.g., \cite{BSW}--\cite{Casa}) as well as the most recent 
predictions obtained using light-cone QCD sum rules \cite{Ball}.
Despite the fact that nonfactorizable corrections are not fully 
controlled theoretically, the estimate (\ref{est}) suggests that 
the SU(3)-breaking corrections in (\ref{dEW}) are small. More 
importantly, such effects cannot induce a sizable strong-interaction 
phase $\omega$. Since $\bar Q_+$ and $\bar Q_-$ are local operators 
whose matrix elements are taken between the same isospin eigenstates, 
it is very unlikely that the strong-interaction phases $\phi_{3/2}^+$ 
and $\phi_{3/2}^-$ could differ by a large amount. If we assume that 
these phases differ by at most $20^\circ$, and that the magnitude of 
$\kappa$ is as large as 12\% (corresponding to twice the central 
value obtained using factorization), we find that 
$|\omega|<2.7^\circ$. Even for a phase difference 
$\Delta\varphi_{3/2}\simeq |\phi_{3/2}^+ - \phi_{3/2}^-|=90^\circ$, 
which seems totally unrealistic, the phase $|\omega|$ would not 
exceed $7^\circ$. It is therefore a safe approximation to work with 
the real value \cite{us}
\begin{equation}
   \delta_{\rm EW}\equiv (1-\kappa)\,\delta_+ = 0.64\pm 0.15 \,,
\end{equation}
where to be conservative we have added linearly the uncertainties 
in the values of $\kappa$ and $\delta_+$. We believe the error 
quoted above is large enough to cover possible small contributions 
from a nonzero phase difference $\Delta\varphi_{3/2}$ or deviations 
from the factorization approximation. For completeness, we note that 
our general results for the structure of the electroweak penguin
contributions to the isospin amplitude $A_{3/2}$, including the
pattern of SU(3)-breaking effects, are in full accord with model
estimates by Deshpande and He \cite{DeHe}. Generalizations of our 
results to the case of $B\to\pi\pi$, $K\bar K$ decays and the 
corresponding $B_s$ decays are possible using SU(3) symmetry, as 
discussed in \cite{Pirjol,Agas}. 

In the last step, we define $K_{3/2}^+\,e^{i\phi_{3/2}^+}
+ K_{3/2}^-\,e^{i\phi_{3/2}^-}\equiv |P|\,\varepsilon_{3/2}\,
e^{i\phi_{3/2}}$, so that \cite{us}
\begin{equation}
   \frac{-3 A_{3/2}}{|P|} = \varepsilon_{3/2}\,e^{i\phi_{3/2}}
   (e^{i\gamma}-\delta_{\rm EW}) \,.
\label{A32simple}
\end{equation}
The complex quantity $q\,e^{i\omega}$ in our general 
parametrization in (\ref{ampl2}) is now replaced with the real
parameter $\delta_{\rm EW}$, whose numerical value is known with
reasonable accuracy. The fact that the strong-interaction phase 
$\omega$ can be neglected was overlooked by Buras and Fleischer, who 
considered values as large as $|\omega|=45^\circ$ and therefore
asigned a larger hadronic uncertainty to the isospin amplitude 
$A_{3/2}$ \cite{BFnew}.

In the SU(3) limit, the product $|P|\,\varepsilon_{3/2}$ is 
determined by the decay amplitude for the process 
$B^\pm\to\pi^\pm\pi^0$ through the relation
\begin{equation}
   |P|\,\varepsilon_{3/2} = \sqrt 2\,\frac{R_{\rm SU(3)}}{R_{\rm EW}}
   \tan\theta_C\,|{\mathcal A}(B^\pm\to\pi^\pm\pi^0)| \,,
\label{SU3rel}
\end{equation}
where\footnote{We disagree with the result for this correction 
presented in \protect\cite{BFnew}.} 
\begin{equation}
   R_{\rm EW} = \left| e^{i\gamma} - \frac{V_{td}}{V_{ud}}\,
   \frac{V_{us}}{V_{ts}}\,\delta_{\rm EW} \right| \simeq \left| 
   1 - \lambda^2 R_t\,\delta_{\rm EW}\,e^{-i\alpha} \right|
\end{equation}
is a tiny correction arising from the very small electroweak penguin 
contributions to the decays $B^\pm\to\pi^\pm\pi^0$. Here 
$\lambda=\sin\theta_C\approx 0.22$ and 
$R_t=[(1-\rho)^2+\eta^2]^{1/2}\sim 1$ are Wolfenstein parameters, and 
$\alpha$ is another angle of the unitarity triangle, whose preferred 
value is close to $90^\circ$ \cite{Jonnew}. It follows that the 
deviation of $R_{\rm EW}$ from 1 is of order 1--2\%, and it is thus a 
safe approximation to set $R_{\rm EW}=1$. More important are 
SU(3)-breaking corrections, which can be included in (\ref{SU3rel}) 
in the factorization approximation, leading to
\begin{equation}
   R_{\rm SU(3)} \simeq \frac{a_1}{a_1+a_2}\,\frac{f_K}{f_\pi}
   + \frac{a_2}{a_1+a_2}\,
   \frac{F_0^{B\to K}(m_\pi^2)}{F_0^{B\to\pi}(m_\pi^2)}
   \simeq \frac{f_K}{f_\pi} \approx 1.2 \,,
\label{RSU3}
\end{equation}
where we have neglected a tiny difference in the phase space for 
the two decays. Relation (\ref{SU3rel}) can be used to 
determine the parameter $\bar\varepsilon_{3/2}$ introduced in
(\ref{bar32}), which coincides with $\varepsilon_{3/2}$ up to
terms of $O(\varepsilon_a)$. To this end, we note that the 
CP-averaged branching ratio for the decays 
$B^\pm\to\pi^\pm K^0$ is given by 
\begin{eqnarray}
   \mbox{Br}(B^\pm\to\pi^\pm K^0) &\equiv& \frac 12 \Big[
    \mbox{Br}(B^+\to\pi^+ K^0) + \mbox{Br}(B^-\to\pi^-\bar K^0)
    \Big] \nonumber\\
   &=& |P|^2\,\Big( 1 - 2\varepsilon_a\cos\eta\cos\gamma
    + \varepsilon_a^2 \Big) \,.
\label{BR1}
\end{eqnarray}
Combining this result with (\ref{SU3rel}) we 
obtain relation (\ref{epsexp}), which expresses 
$\bar\varepsilon_{3/2}$ in terms of CP-averaged branching ratios.
Using preliminary data reported by the CLEO Collaboration \cite{CLEO}
combined with some theoretical guidance based on factorization, one 
finds $\bar\varepsilon_{3/2}=0.24\pm 0.06$ \cite{us}.

To summarize, besides the parameter $\delta_{\rm EW}$ controlling
electroweak penguin contributions also the normalization
of the amplitude $A_{3/2}$ is known from theory, albeit with some 
uncertainty related to nonfactorizable SU(3)-breaking effects. 
The only remaining unknown hadronic parameter in (\ref{A32simple}) 
is the strong-interaction phase $\phi_{3/2}$. The 
various constraints on the structure of the isospin amplitude 
$A_{3/2}$ discussed here constitute the main theoretical 
simplification of $B\to\pi K$ decays, i.e., the only simplification
rooted on first principles of QCD.

\subsection{Structure of the amplitude combination
\boldmath$B_{1/2}+A_{1/2}+A_{3/2}$\unboldmath}

The above result for the isospin amplitude $A_{3/2}$ helps 
understanding better the structure of the sum of amplitudes 
introduced in (\ref{ampl1}). To this end, we introduce the following 
exact parametrization:
\begin{equation}
   B_{1/2} + A_{1/2} + A_{3/2} = |P| \left[ e^{i\pi} e^{i\phi_P}
   - \frac{\varepsilon_{3/2}}{3}\,e^{i\gamma}
   \left( e^{i\phi_{3/2}} - \xi e^{i\phi_{1/2}} \right) \right] \,,
\label{rewrite}
\end{equation}
where we have made explicit the contribution proportional to the 
weak phase $e^{i\gamma}$ contained in $A_{3/2}$. From a comparison 
with the parametrization in (\ref{ampl1}) it follows that
\begin{equation}
   \varepsilon_a\,e^{i\eta} = \frac{\varepsilon_{3/2}}{3}\, 
   e^{i\phi} \left( \xi\,e^{i\Delta} - 1 \right) \,,
\label{repara}
\end{equation} 
where $\phi=\phi_{3/2}-\phi_P$ and $\Delta=\phi_{1/2}-\phi_{3/2}$. 
Of course, this is just a simple reparametrization. However, the
intuitive expectation that $\varepsilon_a$ is small, because this 
terms receives contributions only from the penguin $(P_u-P_t)$ and 
from annihilation topologies, now becomes equivalent to saying 
that $\xi\,e^{i\Delta}$ is close to 1, so as to allow for a 
cancelation between the contributions corresponding to final-state 
isospin $I=\frac 12$ and $I=\frac 32$ in (\ref{repara}). But this 
can only happen if there are no sizable final-state interactions. 
The limit of elastic final-state interactions can be recovered
from (\ref{repara}) by setting $\xi=1$, in which case we reproduce 
results derived previously in \cite{Ge97,Ne97}. Because of the 
large energy release in $B\to\pi K$ decays, however, one expects 
inelastic rescattering contributions to be important as well 
\cite{Fa97,Dono}. They would lead to a value $\xi\ne 1$. 

From (\ref{repara}) it follows that
\begin{equation}
   \varepsilon_a = \frac{\varepsilon_{3/2}}{3}
   \sqrt{1 - 2\xi\cos\Delta + \xi^2}
   = \frac{2\sqrt\xi}{3}\,\varepsilon_{3/2}
   \sqrt{ \left( \frac{1-\xi}{2\sqrt\xi} \right)^2
          + \sin^2\!\frac{\Delta}{2} } \,,
\end{equation}
where without loss of generality we define $\varepsilon_a$ to be
positive. Clearly, $\varepsilon_a\ll\varepsilon_{3/2}$ provided
the phase difference $\Delta$ is small and the parameter $\xi$ 
close to 1. There are good physics reasons to believe that 
both of these requirements may be satisfied. In the rest frame
of the $B$ meson, the two light particles produced in $B\to\pi K$ 
decays have large energies and opposite momenta. Hence, by the 
colour-transparency argument \cite{Bj} their final-state 
interactions are expected to be suppressed unless there are 
close-by resonances, such as charm--anticharm intermediate states 
($D\bar D_s$, $J/\psi\,K$, etc.). However, these contributions 
could only result from the charm penguin \cite{charm1,charming} 
and are thus included in the term $|P|\,e^{i\phi_P}$ in 
(\ref{ampl1}). As a consequence, the phase difference 
$\phi=\phi_{3/2}-\phi_P$ could quite conceivably be sizable. On 
the other hand, the strong phases $\phi_{3/2}$ and $\phi_{1/2}$ 
in (\ref{rewrite}) refer to the matrix elements of local 
four-quark operators of the type $\bar b s\bar u u$ and differ 
only in the isospin of the final state. We believe it is 
realistic to assume that $|\Delta|=|\phi_{1/2}-\phi_{3/2}|<
45^\circ$. Likewise, if the parameter $\xi$ were very different 
from 1 this would correspond to a gross failure of the 
generalized factorization hypothesis (even in decays into isospin 
eigenstates), which works so well in the global analysis of 
hadronic two-body decays of $B$ mesons \cite{Stech}. In view of 
this empirical fact, we think it is reasonable to assume that 
$0.5<\xi<1.5$. With this set of parameters, we find that 
$\varepsilon_a<0.35\varepsilon_{3/2}<0.1$. Thus, we expect that 
the rescattering effects parametrized by $\varepsilon_a$ are 
rather small.

A constraint on the parameter $\varepsilon_a$ can be derived 
assuming $U$-spin invariance of the strong interactions, which 
relates the decay amplitudes for the processes $B^\pm\to\pi^\pm K^0$ 
and $B^\pm\to K^\pm\bar K^0$ up to the substitution 
\cite{Fa97,Robert,He}
\begin{equation}
   \lambda_u \to V_{ub}^* V_{ud} \simeq \frac{\lambda_u}{\lambda}
   \,, \qquad
   \lambda_c \to V_{cb}^* V_{cd} \simeq - \lambda\,\lambda_c \,,
\end{equation}
where $\lambda\approx 0.22$ is the Wolfenstein parameter. 
Neglecting SU(3)-breaking corrections, the CP-averaged 
branching ratio for the decays $B^\pm\to K^\pm\bar K^0$ is then 
given by
\begin{eqnarray}
   \mbox{Br}(B^\pm\to K^\pm\bar K^0) &\equiv& \frac 12 \Big[
    \mbox{Br}(B^+\to K^+\bar K^0) + \mbox{Br}(B^-\to K^- K^0)
    \Big] \nonumber\\
   &=& |P|^2\,\Big[ \lambda^2 + 2\varepsilon_a\cos\eta\cos\gamma
    + (\varepsilon_a/\lambda)^2 \Big] \,,
\end{eqnarray}
which should be compared with the corresponding result for the
decays $B^\pm\to\pi^\pm K^0$ given in (\ref{BR1}). The enhancement 
(suppression) of the subleading (leading) terms by powers of 
$\lambda$ implies potentially large 
rescattering effects and a large direct CP asymmetry in 
$B^\pm\to K^\pm\bar K^0$ decays. In particular, comparing the
expressions for the direct CP asymmetries,
\begin{eqnarray}
   A_{\rm CP}(\pi^+ K^0) &\equiv&
    \frac{\mbox{Br}(B^+\to\pi^+ K^0)-\mbox{Br}(B^-\to\pi^-\bar K^0)}
         {\mbox{Br}(B^+\to\pi^+ K^0)+\mbox{Br}(B^-\to\pi^-\bar K^0)}    
    = \frac{2\varepsilon_a\sin\eta\sin\gamma}
           {1-2\varepsilon_a\cos\eta\cos\gamma+\varepsilon_a^2} \,,
    \nonumber\\
   A_{\rm CP}(K^+\bar K^0)
   &=& - \frac{2\varepsilon_a\sin\eta\sin\gamma}
              {\lambda^2 + 2\varepsilon_a\cos\eta\cos\gamma
               + (\varepsilon_a/\lambda)^2} \,,
\label{ACPs}
\end{eqnarray}
one obtains the simple relation \cite{Robert}
\begin{equation}
   - \frac{A_{\rm CP}(K^+\bar K^0)}{A_{\rm CP}(\pi^+ K^0)}
   = \frac{\mbox{Br}(B^\pm\to\pi^\pm K^0)}
          {\mbox{Br}(B^\pm\to K^\pm\bar K^0)} \,.
\end{equation}
In the future, precise measurements of the branching ratio and 
CP asymmetry in $B^\pm\to K^\pm\bar K^0$ decays may thus provide 
valuable information about the role of rescattering contributions
in $B^\pm\to\pi^\pm K^0$ decays. In particular, upper and lower 
bounds on the parameter $\varepsilon_a$ can be derived from
a measurement of the ratio
\begin{equation}
   R_K = \frac{\mbox{Br}(B^\pm\to K^\pm\bar K^0)}
              {\mbox{Br}(B^\pm\to\pi^\pm K^0)}
   = \frac{\lambda^2 + 2\varepsilon_a\cos\eta\cos\gamma
           + (\varepsilon_a/\lambda)^2}
          {1-2\varepsilon_a\cos\eta\cos\gamma+\varepsilon_a^2} \,.
\label{RKdef}
\end{equation}
Using the fact that $R_K$ is minimized (maximized) by setting 
$\cos\eta\cos\gamma=-1$ (+1), we find that
\begin{equation}
   \frac{\lambda(\sqrt{R_K}-\lambda)}{1+\lambda\sqrt{R_K}}
   \le \varepsilon_a \le
   \frac{\lambda(\sqrt{R_K}+\lambda)}{1-\lambda\sqrt{R_K}} \,.
\end{equation}
This generalizes a relation derived in \cite{Fa97}. Using data
reported by the CLEO Collaboration \cite{CLEO}, one can derive
the upper bound $R_K<0.7$ (at 90\% CL)
implying $\varepsilon_a<0.28$, which is not yet a very powerful 
constraint. However, a measurement of the branching ratio for 
$B^\pm\to K^\pm\bar K^0$ could improve the situation significantly.
For the purpose of illustration, we note that from the preliminary 
results quoted for the observed event rates one may deduce the 
``best fit'' value $R_K\sim 0.15$ (with very large errors!).
Taking this value literally would give the allowed range 
$0.03<\varepsilon_a<0.14$. 

Based on a detailed analysis of individual rescattering 
contributions, Gronau and Rosner have argued that one expects a 
similar pattern of final-state interactions in the decays 
$B^\pm\to K^\pm\bar K^0$ and $B^0\to K^\pm K^\mp$ \cite{GRrescat}. 
One could then use the tighter experimental bound 
$\mbox{Br}(B^0\to K^\pm K^\mp)<2\times 10^{-6}$ to obtain 
$\varepsilon_a<0.16$. However, this is not a model-independent 
result, because the decay amplitudes for $B^0\to K^\pm K^\mp$ are 
not related to those for $B^\pm\to\pi^\pm K^0$ by any symmetry of 
the strong interactions. Nevertheless, this observation may be 
considered a qualitative argument in favour of a small value of 
$\varepsilon_a$.

\subsection{Structure of the amplitude combination 
\boldmath$A_{1/2}+A_{3/2}$\unboldmath}

None of the simplifications we found for the isospin
amplitude $A_{3/2}$ persist for the amplitude $A_{1/2}$. Therefore, 
the sum $A_{1/2}+A_{3/2}$ suffers from larger hadronic
uncertainties than the amplitude $A_{3/2}$ alone. Nevertheless, it
is instructive to study the structure of this combination in more 
detail. In analogy with (\ref{Kdef}), we parametrize the matrix 
elements of the local operators $C_\pm \bar Q_\pm$ between a $B$ 
meson and the $(\pi K)$ isospin state with $I=\frac 12$ by hadronic 
parameters $K_{1/2}^\pm\,e^{i\phi_{1/2}^\pm}$, so that
\begin{equation}
   -3 A_{1/2} = K_{1/2}^+\,e^{i\phi_{1/2}^+} (e^{i\gamma}-\delta_+)
   + K_{1/2}^-\,e^{i\phi_{1/2}^-} (e^{i\gamma}+\delta_+) \,.
\end{equation}
Next, we define parameters $\varepsilon'$ and $r$ by
\begin{equation}
   \frac{\varepsilon'}{2}\,(1\pm r)
   \equiv \frac{2}{3|P|} \left( K_{1/2}^\pm + K_{3/2}^\pm \right) \,.
\end{equation} 
This general definition is motivated by the factorization 
approximation, which predicts that $r\simeq a_2/a_1=0.21\pm 0.05$ is 
the phenomenological colour-suppression factor \cite{Stech}, and 
\begin{equation}
   \frac{\varepsilon'}{\varepsilon_{3/2}}
   \simeq \frac{a_1 A_K}{a_1 A_K + a_2 A_\pi} = 0.84\pm 0.04 \,.
\label{epspr}
\end{equation}
With the help of these definitions, we obtain
\begin{eqnarray}
   \frac{-2(A_{1/2} + A_{3/2})}{|P|}
   &\simeq& \frac{\varepsilon'}{2} \left[ (1+r)\,
    e^{i\phi_{1/2}^+} (e^{i\gamma} - \delta_+) + (1-r)\,
    e^{i\phi_{1/2}^-} (e^{i\gamma} + \delta_+) \right] \nonumber\\
   &&\mbox{}+ \frac{2\varepsilon_{3/2}}{3} \left( e^{i\phi_{3/2}}
    - e^{i\phi_{1/2}^+} \right) (e^{i\gamma} - \delta_+) \,,
\label{newdef}
\end{eqnarray}
where we have neglected some small, SU(3)-breaking corrections to 
the second term. Nevertheless, the above relation can be considered  
a general parametrization of the sum $A_{1/2}+A_{3/2}$, since it 
still contains two undetermined phases $\phi_{1/2}^\pm$ and 
magnitudes $\varepsilon'$ and $r$. 

With the explicit result (\ref{newdef}) at hand, it is a simple
exercise to derive expressions for the quantities entering the 
parametrization in (\ref{ampl2}). We find
\begin{eqnarray}
   \varepsilon_T\,e^{i\phi_T}
   &=& \frac{\varepsilon'}{2}\,e^{i\phi_{1/2}^+} \left[
    (e^{i\Delta\phi_{1/2}}+1) + r(e^{i\Delta\phi_{1/2}}-1) \right]
    + \frac{2\varepsilon_{3/2}}{3} \left( e^{i\phi_{3/2}}
    - e^{i\phi_{1/2}^+} \right) \,, \nonumber\\
   q_C\,e^{i\omega_C} &=& \delta_+\,
    \frac{r(e^{i\Delta\phi_{1/2}}+1) + (e^{i\Delta\phi_{1/2}}-1)
          + \displaystyle\frac{4\varepsilon_{3/2}}{3\varepsilon'}
          \left[ e^{i(\phi_{3/2}-\phi_{1/2}^+)} - 1 \right]} 
         {(e^{i\Delta\phi_{1/2}}+1) + r(e^{i\Delta\phi_{1/2}}-1)
          + \displaystyle\frac{4\varepsilon_{3/2}}{3\varepsilon'} 
          \left[ e^{i(\phi_{3/2}-\phi_{1/2}^+)} - 1 \right]} \,,
\end{eqnarray}
where $\Delta\phi_{1/2}=\phi_{1/2}^- - \phi_{1/2}^+$. This result,
although rather complicated, exhibits in a transparent way the 
structure of possible rescattering effects. In particular, it is 
evident that the assumption of ``colour suppression'' of the 
electroweak penguin contribution, i.e., the statement that 
$q_C=O(r)$ \cite{FM,Robert,BFnew,GR97}, relies on the smallness 
of the strong-interaction phase differences between the various 
terms. More specifically, this assumption would only be justified 
if
\begin{equation}
   |\Delta\phi_{1/2}| < 2 r ~\widehat{=}~ 25^\circ \,, \qquad
   |\phi_{3/2}-\phi_{1/2}^+| < \frac{3r}{2}\,
   \frac{\varepsilon'}{\varepsilon_{3/2}}
   ~\widehat{=}~ 15^\circ \,.
\end{equation}
We believe that, whereas the first relation may be a reasonable 
working hypothesis, the second one constitues a strong constraint 
on the strong-interaction phases, which cannot be justified in a 
model-independent way. As a simple but not unrealistic model we 
may thus consider the approximate relations obtained by setting 
$\Delta\phi_{1/2}=0$ , which have been derived previously in 
\cite{Ne97}:
\begin{eqnarray}
   \varepsilon_T\,e^{i\phi_T}
   &\simeq& \varepsilon'\,e^{i\phi_{1/2}^+} 
    + \frac{2\varepsilon_{3/2}}{3} \left( e^{i\phi_{3/2}}
    - e^{i\phi_{1/2}^+} \right) \,, \nonumber\\
   q_C\,e^{i\omega_C} &\simeq& \delta_+\,
    \frac{r + \displaystyle\frac{2\varepsilon_{3/2}}{3\varepsilon'}
           \left[ e^{i(\phi_{3/2}-\phi_{1/2}^+)} - 1 \right]} 
         {1 + \displaystyle\frac{2\varepsilon_{3/2}}{3\varepsilon'} 
          \left[ e^{i(\phi_{3/2}-\phi_{1/2}^+)} - 1 \right]} \,.
\label{easy}
\end{eqnarray}
The fact that in the case of a sizable phase difference between the
$I=\frac 12$ and $I=\frac 32$ isospin amplitudes the electroweak
penguin contribution may no longer be as small as $O(r)$ has been 
stressed in \cite{Ne97} but was overlooked in \cite{Robert,BFnew}. 
Likewise, there is some uncertainty in the value of the parameter 
$\varepsilon_T$, which in the topological amplitude approach 
corresponds to the ratio $|T-A|/|P|$ \cite{ampl}. Unlike the 
parameter $\varepsilon_{3/2}$, the quantities $\varepsilon'$ and 
$r$ cannot be determined experimentally using SU(3) symmetry 
relations. But even if we assume that the factorization result 
(\ref{epspr}) is valid and take $\varepsilon_{3/2}=0.24$ and 
$\varepsilon'=0.20$ as fixed, we still obtain 
$0.12<\varepsilon_T<0.20$ depending on the value of the phase 
difference $(\phi_{3/2}-\phi_{1/2}^+)$. Note that from the 
approximate expression (\ref{easy}) it follows that
$\varepsilon_T<\varepsilon'$ provided that 
$\varepsilon'/\varepsilon_{3/2}>2/3$, as indicated by the
factorization result. This observation may explain why previous
authors find the value $\varepsilon'=0.15\pm 0.05$ \cite{GR98r}, 
which tends to be somewhat smaller than the factorization 
prediction $\varepsilon'\approx 0.20$. 

\subsection{Numerical results}
\label{subsec:numerics}

Before turning to phenomenological applications of our results in 
the next section, it is instructive to consider some numerical 
results obtained using the above parametrizations. Since our main 
concern in this paper is to study rescattering effects, we will keep 
$\varepsilon_{3/2}=0.24$ fixed and assume that 
$\varepsilon'/\varepsilon_{3/2}=0.84\pm 0.04$ and $r=0.21\pm 0.05$ 
as predicted by factorization. Also, we shall use the factorization 
result for the parameter $\kappa$ in (\ref{est}).

\FIGURE{\epsfig{file=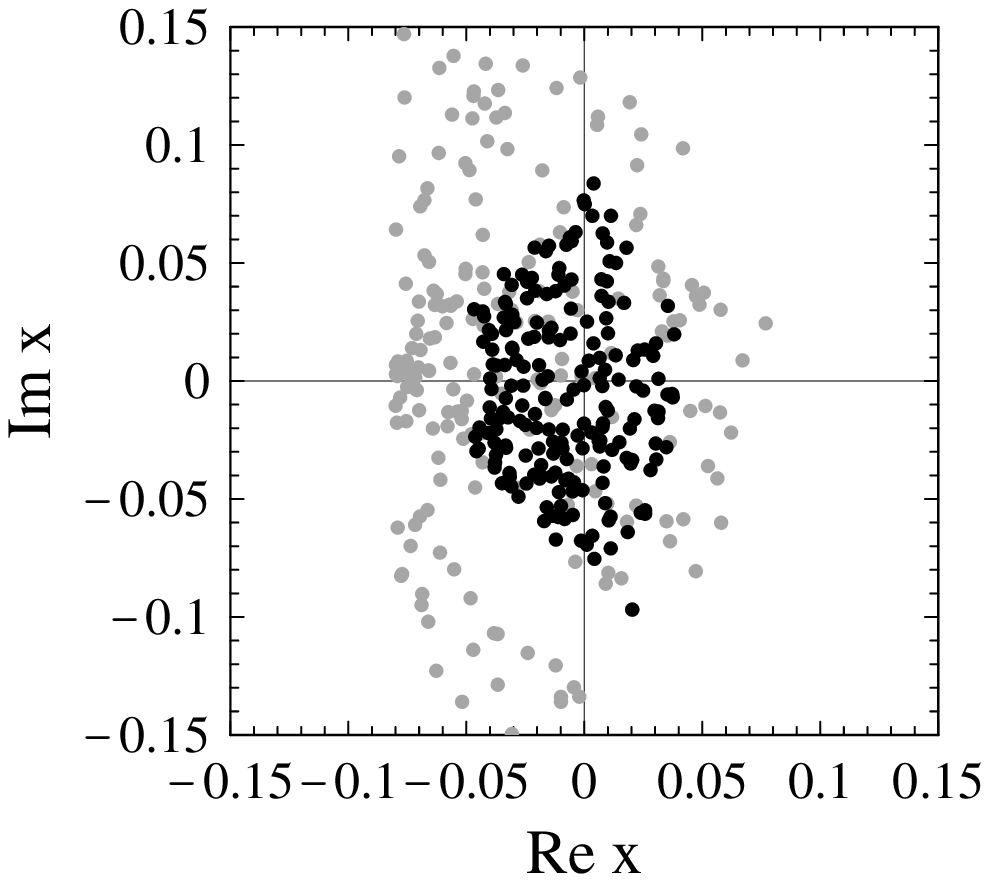,width=7.2cm}
\epsfig{file=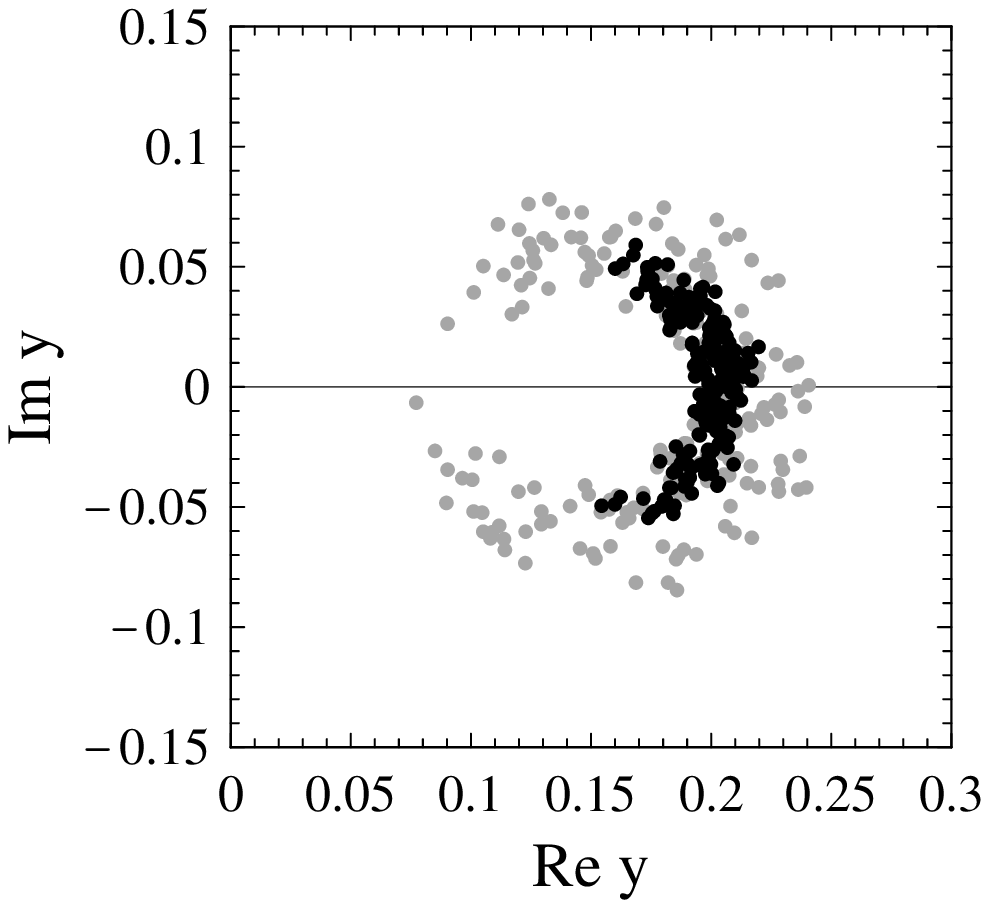,width=7.2cm} 
\caption{Real and imaginary parts of the quantities 
$x=\varepsilon_a\,e^{i(\eta-\phi)}$ (left) and 
$y=\varepsilon_T\,e^{i(\phi_T-\phi_{3/2})}$ (right) for different 
choices of hadronic parameters.}}

\EPSFIGURE{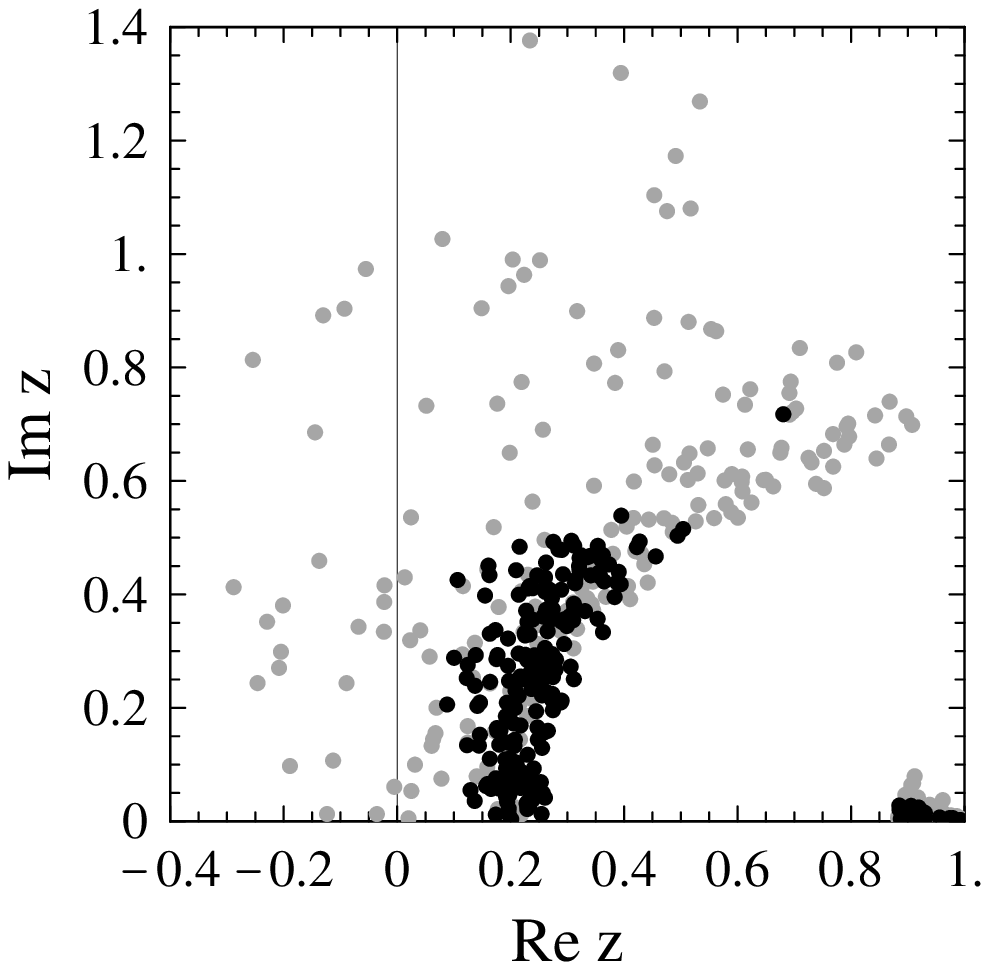,width=7.2cm} 
{Real and imaginary parts of the quantity 
$z=(q_C/\delta_+)\,e^{i\omega_C}$. The accumulation of points in 
the lower-right corner shows the corresponding results for the
quantity $(q/\delta_+)\,e^{i\omega}$. Only points with 
$\mbox{Im}\,z>0$ are shown.}

For the strong-interaction phases we consider two sets of parameter 
choices: one which we believe is realistic and one which we think is 
very conservative. For the realistic set, we require that 
$0.5<\xi<1.5$, $|\phi_{3/2}-\phi_{1/2}^{(+)}|<45^\circ$, and
$|\phi_I^+-\phi_I^-|<20^\circ$ (with $I=\frac 12$ or $\frac 32$). 
For the conservative set, we increase these ranges to $0<\xi<2$, 
$|\phi_{3/2}-\phi_{1/2}^{(+)}|<90^\circ$, 
and $|\phi_I^+-\phi_I^-|<45^\circ$. In our opinion, values outside 
these ranges are quite inconceivable. Note that, for the moment, no 
assumption is made about the relative strong-interaction phases 
of tree and penguin ampltiudes. We choose the various parameters
randomly inside the allowed intervals and present the results
for the quantities $\varepsilon_a\,e^{i\eta}$ in units of 
$e^{i\phi}$, $\varepsilon_T\,e^{i\phi_T}$ in units of 
$e^{i\phi_{3/2}}$, and $q_{(C)}\,e^{i\omega_{(C)}}$  in units of 
$\delta_+$ in the form of scatter plots in Figures~1 and 2. The 
black and the gray points correspond to the realistic and to the 
conservative parameter sets, respectively. The same colour coding 
will be used throughout this work.

The left-hand plot in Figure~1 shows that the parameter 
$\varepsilon_a$ generally 
takes rather small values. For the realistic parameter set we find 
$\varepsilon_a<0.08$, whereas values up tp 0.15 are possible for the 
conservative set. There is no strong correlation between the 
strong-interaction phases $\eta$ and $\phi$. An important 
implication of these observations is that, in general, there will
be a very small difference between the quantities 
$\varepsilon_{3/2}$ and $\bar\varepsilon_{3/2}$ in (\ref{bar32}). We
shall therefore consider the same range of values for the two 
parameters. From the right-hand plot we observe that for realistic 
parameter choices $0.15<\varepsilon_T<0.22$; however, values between 
0.08 and 0.24 are possible for the conservative parameter set. Note 
that there is a rather strong 
correlation between the strong-interaction phases $\phi_T$ and 
$\phi_{3/2}$, which differ by less than $20^\circ$ for the realistic 
parameter set. We will see in Section~\ref{sec:5} that this implies a 
strong correlation between the direct CP asymmetries in the decays 
$B^\pm\to\pi^0 K^\pm$ and $B^0\to\pi^\mp K^\pm$. Figure~2 shows 
that, even for the realistic parameter set, the ratio $q_C/\delta_+$ 
can be substantially larger than the naive expectation of about 0.2. 
Indeed, values as large as 0.7 are possible, and for the conservative 
set the wide range $0<q_C/\delta_+<1.4$ is allowed. Likewise, the 
strong-interaction phase $\omega_C$ can naturally be large and take 
values of up to $75^\circ$ even for the realistic parameter set. 
(Note that, without loss of generality, only points with positive
values of $\omega_{(C)}$ are displayed in the plot. The distribution 
is invariant under a change of the sign of the strong-interaction 
phase.) This is in stark contrast to the case of the quantity 
$q\,e^{i\omega}$ entering the isospin amplitude $A_{3/2}$, where both 
the magnitude $q$ and the phase $\omega$ are determined within very 
small uncertainties, as is evident from the figure.

\section{Hadronic uncertainties in the Fleischer--Mannel bound}
\label{sec:3}

As a first phenomenological application of the results of the 
previous section, we investigate the effects of rescattering 
and electroweak penguin contributions on the Fleischer--Mannel bound 
on $\gamma$ derived from the ratio $R$ defined in (\ref{Rdef}). 
In general, $R\ne 1$ because the parameter $\varepsilon_T$ in 
(\ref{ampl2}) does not vanish. To leading order in the small 
quantities $\varepsilon_i$, we find
\begin{equation}
   R\simeq 1 - 2\varepsilon_T \Big[ \cos\tilde\phi\cos\gamma
   - q_C\cos(\tilde\phi+\omega_C) \Big] + O(\varepsilon_i^2) \,,
\end{equation}
where $\tilde\phi=\phi_T-\phi_P$. Because of the uncertainty in
the values of the hadronic parameters $\varepsilon_T$, $q_C$ and 
$\omega_C$, it is difficult to convert this result into 
a constraint on $\gamma$. Fleischer and Mannel have therefore 
suggested to derive a lower bound on the ratio $R$ by eliminating
the parameter $\varepsilon_T$ from the exact expression for $R$. In 
the limit where $\varepsilon_a$ and $q_C$ are set to zero, this 
yields $R\ge\sin^2\!\gamma$ \cite{FM}. However, this simple result 
must be corrected in the presence of rescattering effects and 
electroweak penguin contributions. The generalization is 
\cite{Robert}
\begin{equation}
   R \ge \frac{1-2q_C\,\varepsilon_a\cos(\omega_C+\eta)
               +q_C^2\,\varepsilon_a^2}
          {(1-2q_C\cos\omega_C\cos\gamma+q_C^2)
           (1-2\varepsilon_a\cos\eta\cos\gamma+\varepsilon_a^2)}\,
   \sin^2\!\gamma \,.                
\label{Rob}
\end{equation}
The most dangerous rescattering effects arise from the terms
involving the electroweak penguin parameter $q_C$. As seen from
Figure~2, even restricting ourselves to the realistic parameter 
set we can have $2q_C\cos\omega_C\approx\delta_+\approx 0.7$ 
and $q_C^2\approx 0.5\,\delta_+^2\approx 0.2$, implying that the 
quadratic term in the denominator by itself can give a 20\% 
correction. The rescattering effects parametrized by $\varepsilon_a$ 
are presumably less important.   

\EPSFIGURE{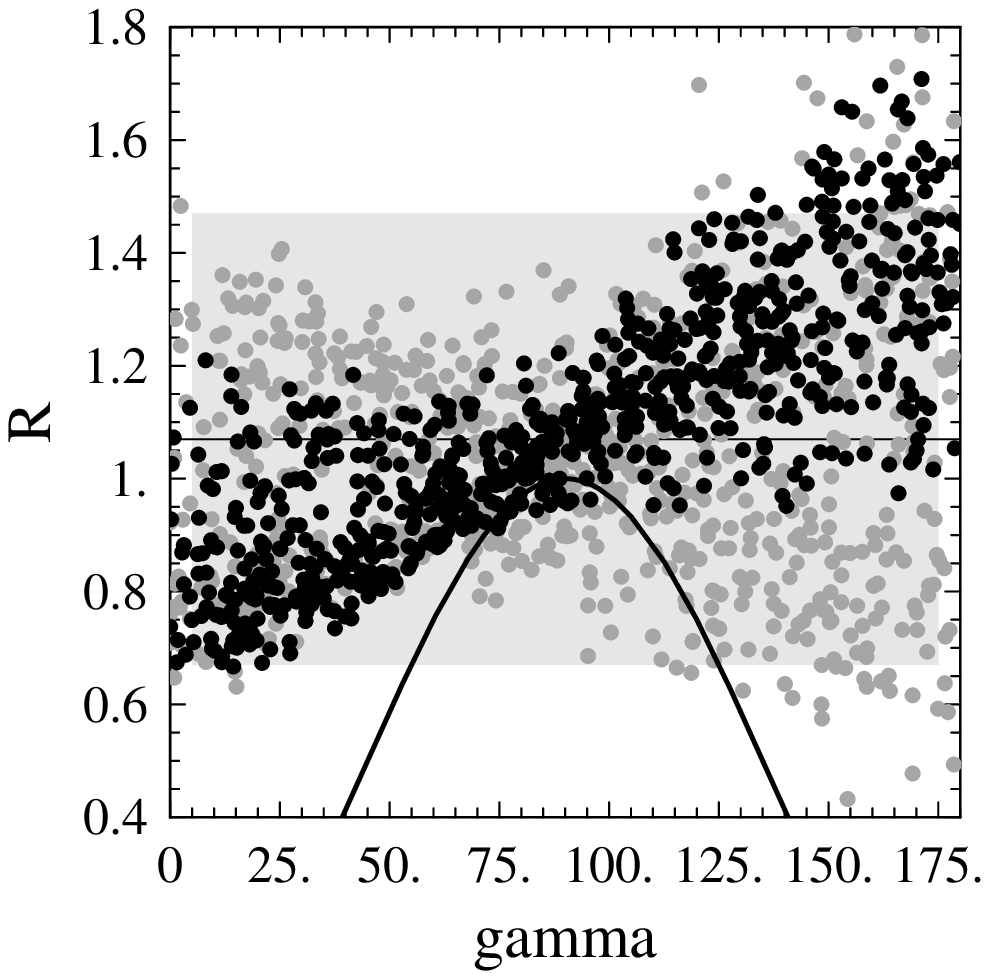,width=7.2cm} 
{Results for the ratio $R$ versus $|\gamma|$ (in degrees) for 
different choices of hadronic parameters. The curve shows the 
minimal value $R_{\rm min}=\sin^2\!\gamma$ corresponding to the 
Fleischer--Mannel bound. The band shows the current experimental 
value of $R$ with its $1\sigma$ variation.}

The results of the numerical analysis are shown in Figure~3. In 
addition to the parameter choices described in 
Section~\ref{subsec:numerics}, we vary $\varepsilon_{3/2}$ and 
$\delta_+$ in the ranges $0.24\pm 0.06$ and $0.68\pm 0.11$, 
respectively. Now also the relative strong-interaction phase 
$\phi$ between the penguin and $I=\frac 32$ tree amplitudes enters. 
We allow values $|\phi|<90^\circ$ for the realistic parameter set, 
and impose no constraint on $\phi$ at all for the conservative 
parameter set. The figure shows that the corrections to the 
Fleischer--Mannel bound are not as large as suggested by the result 
(\ref{Rob}), the reason being that this result is derived allowing 
arbitrary values of $\varepsilon_T$, whereas in our analysis the 
allowed values for this parameter are constrained. However, there 
are sizable violations of the naive bound $R<\sin^2\!\gamma$ for 
$|\gamma|$ in the range between $65^\circ$ and $125^\circ$, which 
includes most of the region $47^\circ<\gamma<105^\circ$ preferred 
by the global analysis of the unitarity triangle \cite{Jonnew}. 
Whereas these violations are numerically small for the realistic 
parameter set, they can become large for the conservative set, 
because then a large value of the phase difference 
$|\phi_{3/2}-\phi_{1/2}^+|$ is allowed \cite{Ne97}. We conclude 
that under conservative assumptions only for values $R<0.8$ a 
constraint on $\gamma$ can be derived

Fleischer has argued that one can improve upon the above analysis 
by extracting some of the unknown hadronic parameters $q_C$, 
$\varepsilon_a$, $\omega_C$ and $\eta$ from measurements of other 
decay processes \cite{Robert}. The idea is to combine information 
on the ratio $R$ with measurements of the direct CP asymmetries in 
the decays $B^0\to\pi^\mp K^\pm$ and $B^\pm\to\pi^\pm K^0$, as well 
as of the ratio $R_K$ defined in (\ref{RKdef}). One can then derive 
a bound on $R$ that depends, besides the electroweak penguin 
parameters $q_C$ and $\omega_C$, only on a combination 
$w=w(\varepsilon_a,\eta)$, which can be determined up to a two-fold 
ambiguity assuming SU(3) flavour symmetry. Besides the fact that 
this approach relies on SU(3) symmetry and involves significantly 
more experimental input than the original Fleischer--Mannel 
analysis, it does not allow one to eliminate the theoretical 
uncertainty related to the presence of electroweak penguin 
contributions.

\section{Hadronic uncertainties in the \boldmath$R_*$\unboldmath\
bound}
\label{sec:4}

As a second application, we investigate the implications of
recattering effects on the bound on $\cos\gamma$ derived from a 
measurement of the ratio $R_*$ defined in (\ref{Rdef}). In this 
case, the theoretical analysis is cleaner because there is 
model-independent information on the values of the hadronic
parameters $\varepsilon_{3/2}$, $q$ and $\omega$ entering the
parametrization of the isospin amplitude $A_{3/2}$ in 
(\ref{ampl2}). The important point noted in \cite{us} is that
the decay amplitudes for $B^\pm\to\pi^\pm K^0$ 
and $B^\pm\to\pi^0 K^\pm$ differ only in this single isospin
amplitude. Since the overall strength of $A_{3/2}$ is governed
by the parameter $\bar\varepsilon_{3/2}$ and thus can
be determined from experiment without much uncertainty, we
have suggested to derive a bound on $\cos\gamma$ without
eliminating this parameter. In this respect, our strategy is 
different from the Fleischer--Mannel analysis.  

The exact theoretical expression for the inverse of the ratio 
$R_*$ is given by
\begin{eqnarray}
   R_*^{-1} &=& 1 + 2\bar\varepsilon_{3/2}\,
    \frac{\cos\phi\,(\delta_{\rm EW}-\cos\gamma)
          + \varepsilon_a\cos(\phi-\eta)(1-\delta_{\rm EW}\cos\gamma)}
         {\sqrt{1-2\varepsilon_a\cos\eta\cos\gamma+\varepsilon_a^2}}
    \nonumber\\
   &&\mbox{}+ \bar\varepsilon_{3/2}^2
    (1-2\delta_{\rm EW}\cos\gamma+\delta_{\rm EW}^2) \,,
\label{R*}
\end{eqnarray}
where $\bar\varepsilon_{3/2}$ has been defined in (\ref{bar32}). 
Relevant for the bound on $\cos\gamma$ is the maximal value $R_*^{-1}$
can take for fixed $\gamma$. In \cite{us}, we have worked to linear 
order in the parameters $\varepsilon_i$, so that terms proportional to 
$\varepsilon_a$ could be neglected. Here, we shall generalize the 
discussion and keep all terms exactly. Varying the strong-interaction
phases $\phi$ and $\eta$ independently, we find that the maximum value 
of $R_*^{-1}$ is given by
\begin{equation}
   R_*^{-1} \le 1 + 2\bar\varepsilon_{3/2}\,
   \frac{|\delta_{\rm EW}-\cos\gamma \pm \varepsilon_a
         (1-\delta_{\rm EW}\cos\gamma)|}
        {\sqrt{1 \mp 2\varepsilon_a\cos\gamma+\varepsilon_a^2}}
   + \bar\varepsilon_{3/2}^2
   (1-2\delta_{\rm EW}\cos\gamma+\delta_{\rm EW}^2) \,,
\label{exact}
\end{equation}
where the upper (lower) signs apply if $\cos\gamma<c_0$
($\cos\gamma>c_0$) with
\begin{equation}
   c_0 = \frac{(1+\varepsilon_a^2)\,\delta_{\rm EW}}
                 {1+\varepsilon_a^2\,\delta_{\rm EW}^2}
   \simeq \delta_{\rm EW} \,.
\label{c0}
\end{equation}
Keeping all terms in $\bar\varepsilon_{3/2}$ exactly, but working 
to linear order in $\varepsilon_a$, we find the simpler result
\begin{equation}
   R_*^{-1} \le \Big( 1 + \bar\varepsilon_{3/2}\,
   |\delta_{\rm EW}-\cos\gamma| \Big)^2 + \bar\varepsilon_{3/2}
   (\bar\varepsilon_{3/2} + 2\varepsilon_a) \sin^2\!\gamma
   + O(\bar\varepsilon_{3/2}\,\varepsilon_a^2) \,.
\label{Rstmax}
\end{equation}
The higher-order terms omitted here are of order 1\% and thus 
negligible. The annihilation contribution $\varepsilon_a$ enters 
this result in a very transparent way: increasing $\varepsilon_a$ 
increases the maximal value of $R_*^{-1}$ and therefore weakens 
the bound on $\cos\gamma$.

In \cite{us}, we have introduced the quantity $\Delta_*$ by
writing $R_*=(1-\Delta_*)^2$, so that $\Delta_*=1-\sqrt{R_*}$ obeys
the bound shown in (\ref{ourb}). Note that to first order in 
$\bar\varepsilon_{3/2}$ the rescattering contributions proportional 
to $\varepsilon_a$ do not enter.\footnote{Contrary to what has been 
claimed in \protect\cite{BFnew}, this does not mean that we were 
ignoring rescattering effects altogether. At linear order, these 
effects enter only through the strong-interaction phase difference 
$\phi$, which we kept arbitrary in deriving the bound on 
$\cos\gamma$.}
Armed with the result (\ref{exact}), we can now derive the exact 
expression for the maximal value of the quantity $\Delta_*$, 
corresponding to the minimal value of $R_*$. It is of advantage to 
consider the ratio $\Delta_*/\bar\varepsilon_{3/2}$, the bound for 
which is to first order independent of the parameter 
$\bar\varepsilon_{3/2}$. We recall that this ratio can be determined 
experimentally up to nonfactorizable SU(3)-breaking corrections. Its 
current value is $\Delta_*/\bar\varepsilon_{3/2}=1.33\pm 0.78$.

\FIGURE{\epsfig{file=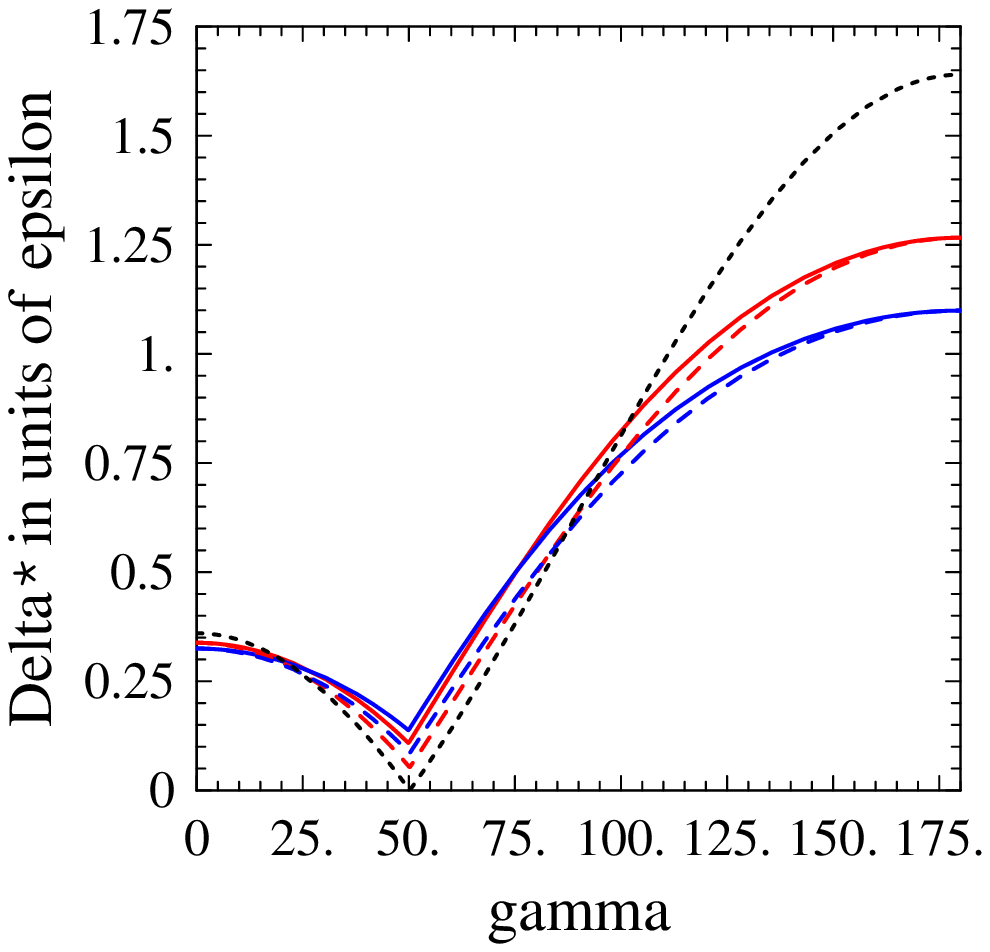,width=7.2cm}
\epsfig{file=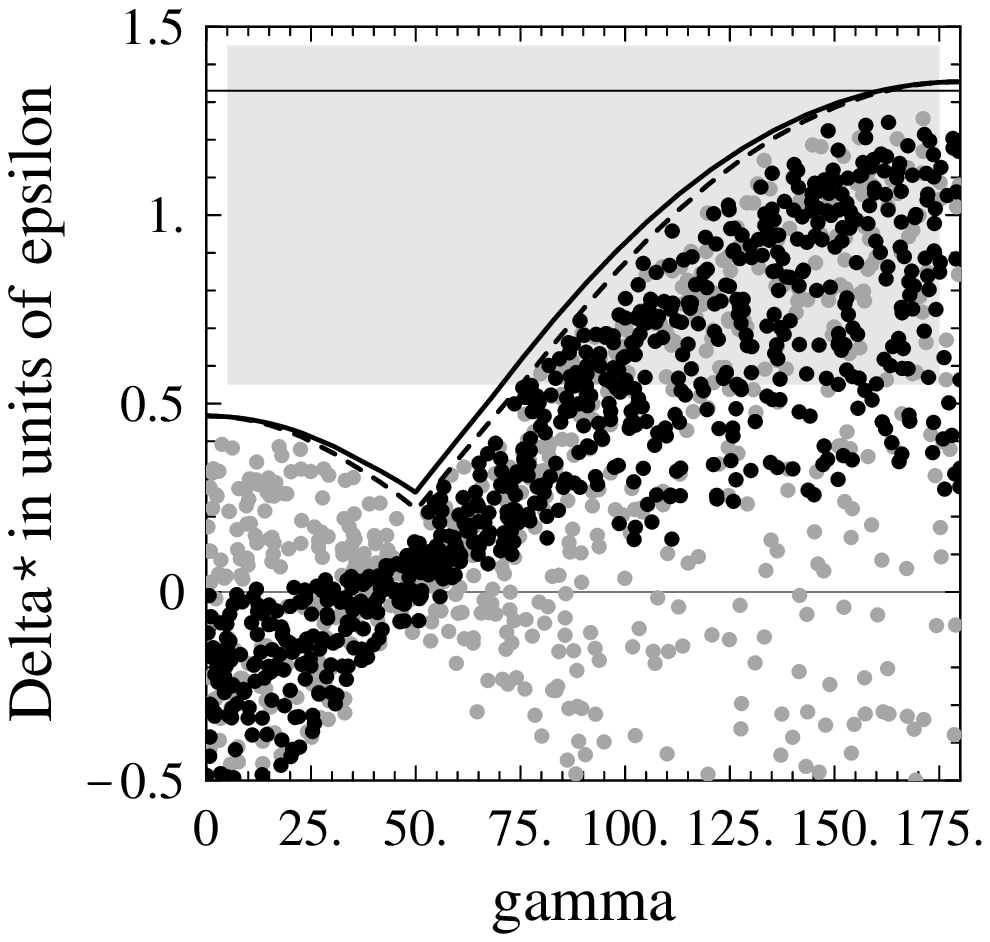,width=7.2cm} 
\caption{Left: Theoretical upper bound on the ratio 
$\Delta_*/\bar\varepsilon_{3/2}$ versus $|\gamma|$ for various 
choices of $\bar\varepsilon_{3/2}$ and $\varepsilon_a$, as explained 
in the text. The dotted line shows the linear bound derived in 
\protect\cite{us}. Right: Results for this ratio obtained for 
different choices of hadronic parameters. The curves show the 
theoretical bound for $\varepsilon_a=0.1$ (solid) and 
$\varepsilon_a=0$ (dashed). The band shows the current experimental 
value with its $1\sigma$ variation.}}

In the left-hand plot in Figure~4, we show the maximal value for the 
ratio $\Delta_*/\bar\varepsilon_{3/2}$ for different values of the 
parameters $\bar\varepsilon_{3/2}$ and $\varepsilon_a$. The upper 
(red) and lower (blue) pairs of curves correspond to 
$\bar\varepsilon_{3/2}=0.18$ and 0.30, respectively, and span the 
allowed range of values for this parameter. For each pair, the 
dashed and solid lines correspond to $\varepsilon_a=0$ and 0.1, 
respectively. To saturate the bound (\ref{exact}) requires to have 
$\eta-\phi=0^\circ$ or $180^\circ$, in which case 
$\varepsilon_a=0.1$ is a conservative upper limit (see Figure~1). 
The dotted curve shows for comparison the linearized result obtained 
by neglecting the higher-order terms in (\ref{ourb}). The parameter
$\delta_{\rm EW}=0.64$ is kept fixed in this plot. As expected, the 
bound on the ratio $\Delta_*/\bar\varepsilon_{3/2}$ is only weakly 
dependent on the values of $\bar\varepsilon_{3/2}$ and 
$\varepsilon_a$. In particular, not much is lost by using the
conservative value $\varepsilon_a=0.1$. Note that for 
values $\Delta_*/\bar\varepsilon_{3/2}>0.8$ the linear bound 
(\ref{ourb}) is conservative, i.e., weaker than the exact bound,
and even for smaller values of $\Delta_*/\bar\varepsilon_{3/2}$ the 
violations of this bound are rather small. Expanding the exact bound 
to next-to-leading order in $\bar\varepsilon_{3/2}$, we obtain
\begin{equation}
   \frac{\Delta_*}{\bar\varepsilon_{3/2}}
   \le |\delta_{\rm EW}-\cos\gamma| - \bar\varepsilon_{3/2} \left[
   \left( \frac{\Delta_*}{\bar\varepsilon_{3/2}} \right)^2
   - \left( \frac12 + \frac{\varepsilon_a}{\bar\varepsilon_{3/2}}
   \right) \sin^2\!\gamma \right] + O(\bar\varepsilon_{3/2}^2) \,,
\end{equation}
showing that $\Delta_*/\bar\varepsilon_{3/2}>(1/2+\varepsilon_a/
\bar\varepsilon_{3/2})^{1/2}$ is a criterion for the validity of the 
linearized bound. This generalizes a condition derived, for the 
special case $\varepsilon_a\ll\bar\varepsilon_{3/2}$, in \cite{us}. 

To obtain a reliable bound on the weak phase $\gamma$, we must 
account for the theoretical uncertainty in the value of the 
electroweak penguin parameter $\delta_{\rm EW}$ in the Standard
Model, which is however straightforward to do by lowering 
(increasing) the value of this parameter used in calculating the
right (left) branch of the curves defining the bound. The solid
line in the right-hand plot in Figure~4 shows the most conservative
bound obtained by using $\varepsilon_a=0.1$ and varying the 
other two parameters in the ranges $0.18<\bar\varepsilon_{3/2}
<0.30$ and $0.49<\delta_{\rm EW}<0.79$. The scatter plot shows
the distribution of values of $\Delta_*/\bar\varepsilon_{3/2}$
obtained by scanning the strong-interaction parameters over the
same ranges as we did for the Fleischer--Mannel case in the 
previous section. The horizontal band shows the current central 
experimental value with its $1\sigma$ variation. Unlike the
Fleischer--Mannel bound, there is no violation of the bound 
(by construction), since all parameters are varried over 
conservative ranges. Indeed, for the points close to the right
branch of the bound $\eta-\phi=0^\circ$, so that according to 
Figure~1 almost all of these points have $\varepsilon_a<0.03$, 
which is smaller than the value we used to obtain the theoretical 
curve. The dashed curve shows the bound for $\varepsilon_a=0$, 
which is seen not to be violated by any point. This shows that 
the rescattering effects parametrized by the quantity 
$\varepsilon_a$ play a very minor role in the bound derived from 
the ratio $R_*$. We conclude that, if the current experimental 
value is confirmed to within one standard deviation, i.e., if 
future measurements find that $\Delta_*/\bar\varepsilon_{3/2}
>0.55$, this would imply the bound $|\gamma|>75^\circ$, which is 
very close to the value of $77^\circ$ obtained in \cite{us}.

\FIGURE{\epsfig{file=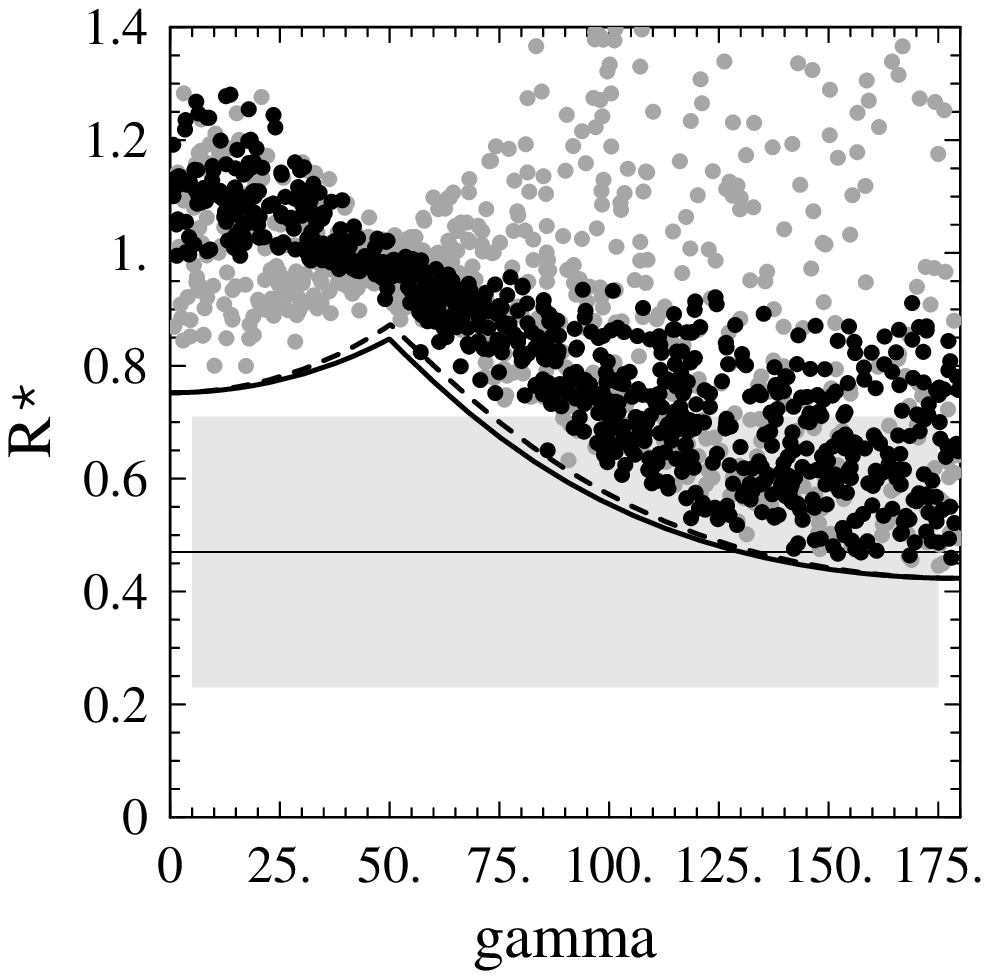,width=7.2cm}
\epsfig{file=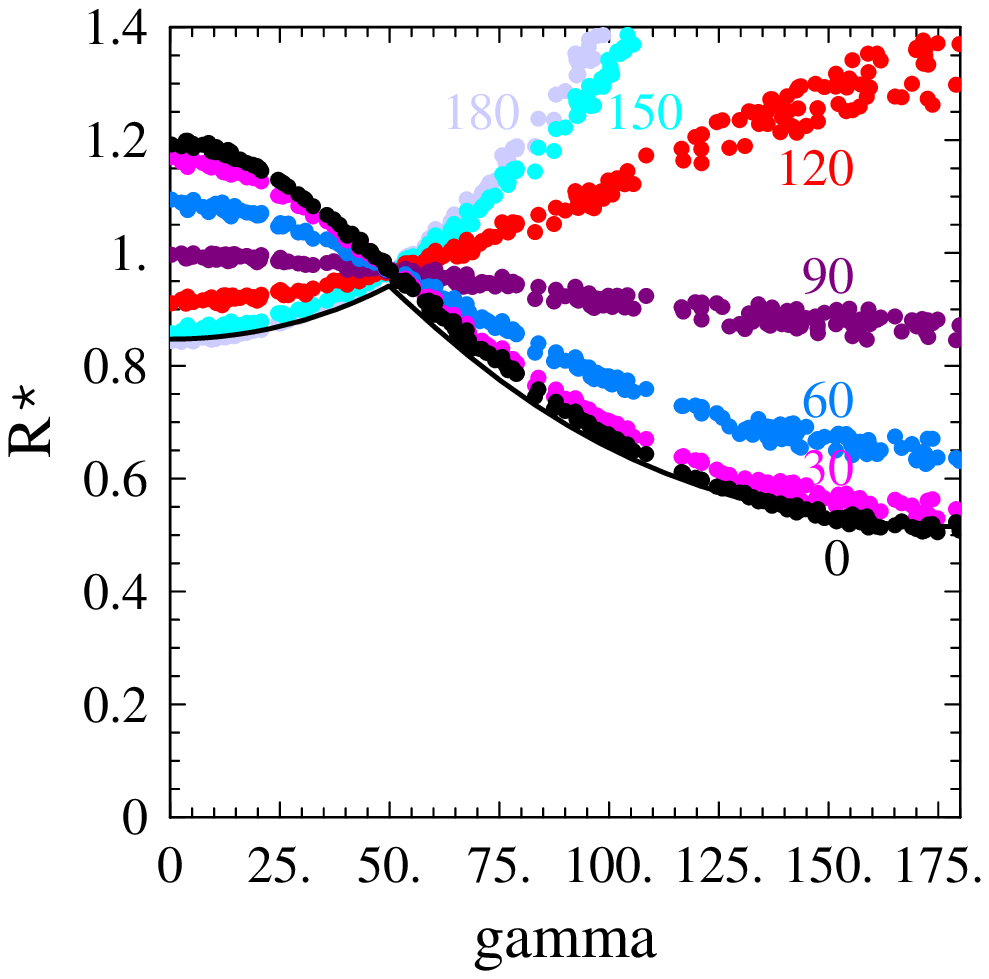,width=7.2cm} 
\caption{Left: Results for the ratio $R_*$ obtained for different 
choices of hadronic parameters. The curves show the theoretical 
bound for $\varepsilon_a=0.1$ (solid) and $\varepsilon_a=0$ 
(dashed). The band shows the current experimental value of $R_*$
with its $1\sigma$ variation. Right: Results for the ratio $R_*$ 
for different values of the strong-interaction phase $|\phi|$, as 
indicated by the numbers.}}

Given that the experimental determination of the parameter
$\bar\varepsilon_{3/2}$ is limited by unknown nonfactorizable
SU(3)-breaking corrections, one may want to be more conservative
and derive a bound directly from the measured ratio $R_*$ rather
than the ratio $\Delta_*/\bar\varepsilon_{3/2}$. In the left-hand 
plot in Figure~5, we show the same distribution as in the 
right-hand plot 
in Figure~4, but now for the ratio $R_*$. The resulting bound on
$\gamma$ is slightly weaker, because now there is a stronger
dependence on the value of $\bar\varepsilon_{3/2}$, which we vary 
as previously between 0.18 and 0.30. If the current value of $R_*$
is confimed to within one standard deviation, i.e., if future 
measurements find that $R_*<0.71$, this would imply the bound 
$|\gamma|>72^\circ$.

Besides providing interesting information on $\gamma$, 
a measurement of $R_*$ or $\Delta_*/\bar\varepsilon_*$ can  
yield information about the strong-interaction phase $\phi$. In the 
right plot in Figure~5, we show the distribution of points obtained 
for fixed values of the strong-interaction phase $|\phi|$ between 
$0^\circ$ and $180^\circ$ in steps of $30^\circ$. For simplicity, 
the parameters 
$\varepsilon_{3/2}=0.24$ and $\delta_{\rm EW}=0.64$ are kept fixed 
in this plot, while all other hadronic parameters are scanned over 
the realistic parameter set. We observe that, independently of 
$\gamma$, a value $R_*<0.8$ requires that $|\phi|<90^\circ$. This 
conclusion remains true if the parameters $\varepsilon_{3/2}$ and 
$\delta_{\rm EW}$ are varied over their allowed ranges. We shall 
study the correlation between the weak phase $\gamma$ and the 
strong phase $\phi$ in more detail in Section~\ref{sec:6}.

\EPSFIGURE{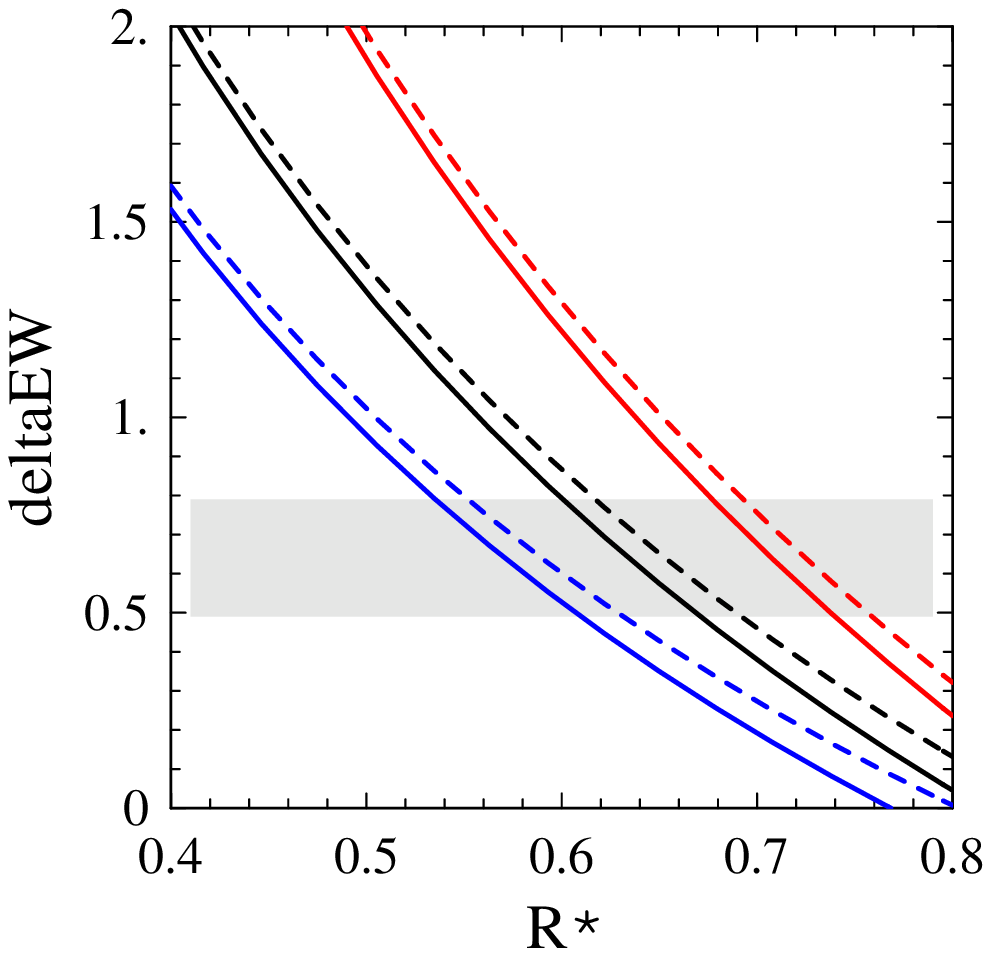,width=7.2cm} 
{Lower bound on the parameter $\delta_{\rm EW}$ as a function of
$R_*$. The upper (red), middle (black) and lower (blue) 
curves correspond to $\bar\varepsilon_{3/2}=0.18$, 0.24 and 0.30, 
respectively. The solid and dashed lines refer to 
$\varepsilon_a=0.1$ and 0. The band shows the prediction for
$\delta_{\rm EW}$ in the Standard Model.}

Finally, we emphasize that a future, precise measurement of the
ratio $R_*$ may also yield a surprise and indicate physics beyond
the Standard Model. The global analysis of the unitarity triangle
requires that $|\gamma|<105^\circ$ \cite{Jonnew}, for which the 
lowest possible value of $R_*$ in the Standard Model is about 0.55. 
If the experimental value would turn out to be less than that, this 
would be strong evidence for New Physics. In particular, in many 
extensions of the Standard Model there would be additional 
contributions to the electroweak penguin parameter 
$\delta_{\rm EW}$ arising, e.g., from penguin and box diagrams 
containing new charged Higgs bo\-sons. This could explain a larger 
value of $R_*$. Indeed, from (\ref{Rstmax}) we can derive the bound
\begin{eqnarray}
   \delta_{\rm EW} \!&\ge&\! 
    \frac{\sqrt{R_*^{-1} - \bar\varepsilon_{3/2}
          (\bar\varepsilon_{3/2}+2\varepsilon_a)
          \sin^2\!\gamma_{\rm max}} - 1}{\bar\varepsilon_{3/2}}
    \nonumber\\
   &&\mbox{}+ \cos\gamma_{\rm max} \,,
\end{eqnarray}
where $\gamma_{\rm max}$ is the maximal value allowed by the global 
analysis (assuming that $\gamma_{\rm max}>\arccos(c_0)\approx 
50^\circ$). In Figure~6, we show this bound for the current value 
$\gamma_{\rm max}=105^\circ$ and three different values of 
$\bar\varepsilon_{3/2}$ as well as two different values of 
$\varepsilon_a$. The gray band shows the allowed range for 
$\delta_{\rm EW}$ in the Standard Model. In the hypothetical 
situation where the current central values $R_*=0.47$ and 
$\bar\varepsilon_{3/2}=0.24$ would be confirmed by more precise 
measurements, we would conclude that the value of $\delta_{\rm EW}$
is at least twice as large as predicted by the Standard Model.

\section{Prospects for direct CP asymmetries and prediction for
the\\ 
\boldmath$B^0\to\pi^0 K^0$\unboldmath\ branching ratio}
\label{sec:5}

\subsection{Decays of charged \boldmath$B$\unboldmath\ mesons}

We will now analyse the potential of the various $B\to\pi K$ decay
modes for showing large direct CP violation, starting with the
decays of charged $B$ mesons. The smallness of the rescattering
effects parametrized by $\varepsilon_a$ (see Figure~1) combined with 
the simplicity of the isospin amplitude $A_{3/2}$ (see 
Section~\ref{subsec:A32}) make these processes particularly clean
from a theoretical point of view.

Explicit expressions for the CP asymmetries in the various decays 
can be derived in a straightforward way starting from the isospin 
decomposition in (\ref{isodec}) and inserting the parametrizations 
for the isospin amplitudes derived in Section~\ref{sec:2}. The 
result for the CP asymmetry in the decays $B^\pm\to\pi^\pm K^0$ has 
already been presented in (\ref{ACPs}). The corresponding expression 
for the decays $B^\pm\to\pi^0 K^\pm$ reads
\begin{equation}
   A_{\rm CP}(\pi^0 K^+) = 2\sin\gamma\,R_*\,
   \frac{\varepsilon_{3/2}\sin\phi+\varepsilon_a\sin\eta
         -\varepsilon_{3/2}\,\varepsilon_a\,\delta_{\rm EW}
          \sin(\phi-\eta)}
        {1-2\varepsilon_a\cos\eta\cos\gamma+\varepsilon_a^2} \,,
\label{ACP2}
\end{equation}
where the theoretical expression for $R_*$ is given in (\ref{R*}), 
and we have not replaced $\varepsilon_{3/2}$ in terms of
$\bar\varepsilon_{3/2}$. Neglecting terms of order $\varepsilon_a$ and
working to first order in $\varepsilon_{3/2}$, we find the estimate 
$A_{\rm CP}(\pi^0 K^+)\simeq 2\varepsilon_{3/2}\sin\gamma
\sin\phi\approx 0.5\sin\gamma\sin\phi$, indicating that 
potentially there could be a very large CP asymmetry in this decay
(note that $\sin\gamma>0.73$ is required by the global analysis of
the unitarity triangle).

\FIGURE{\epsfig{file=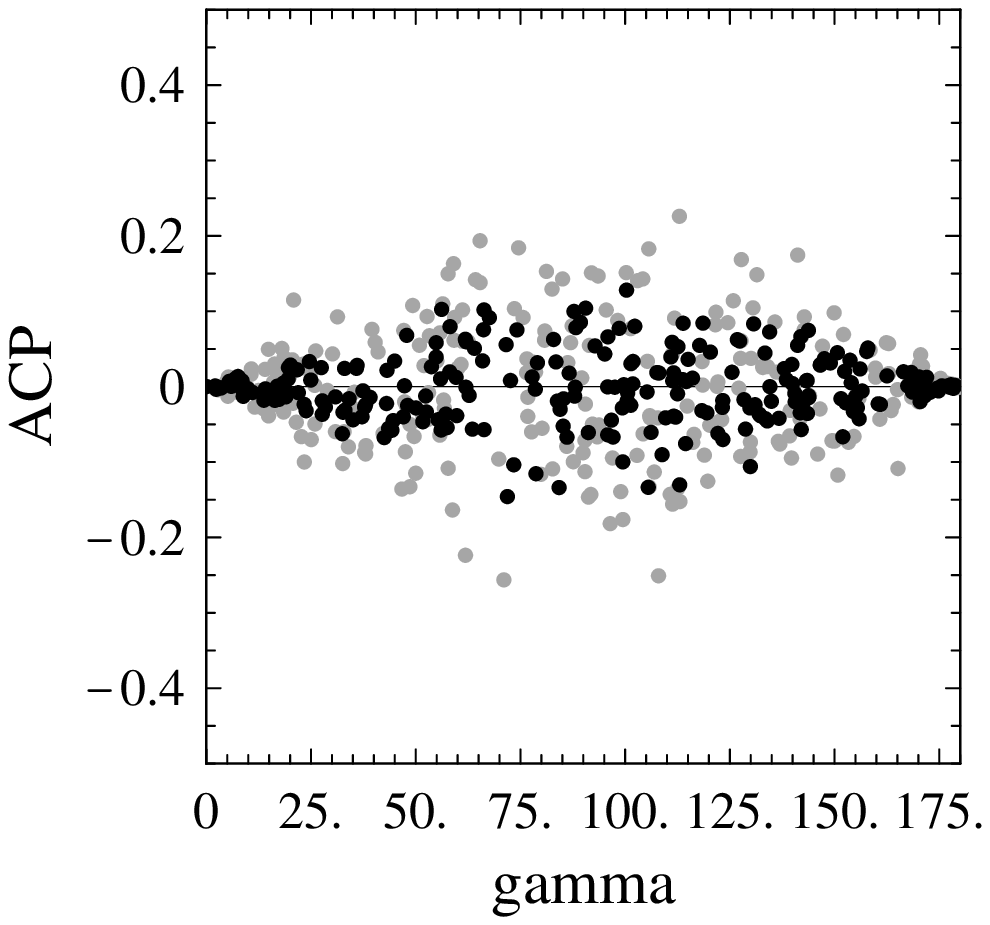,width=7.2cm}
\epsfig{file=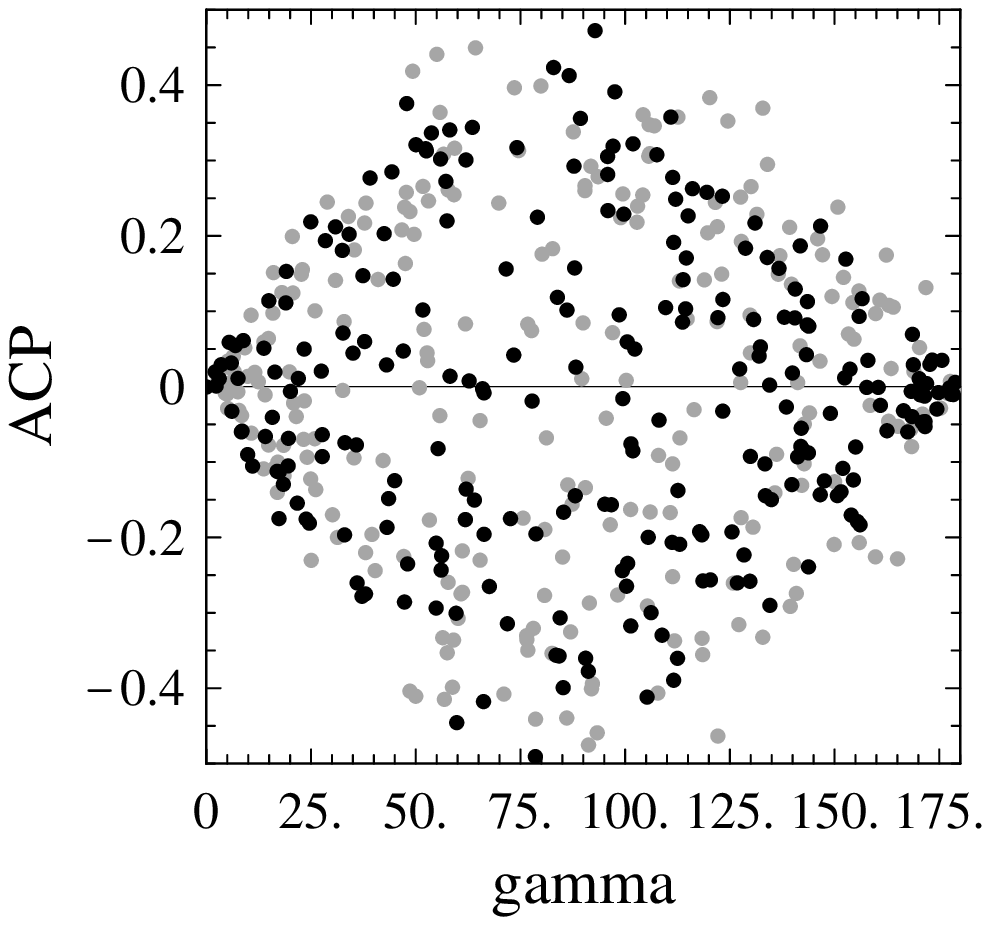,width=7.2cm} 
\caption{Results for the direct CP asymmetries $A_{\rm CP}(\pi^+ K^0)$ 
(left) and $A_{\rm CP}(\pi^0 K^+)$ (right) versus $|\gamma|$.}}

In Figure~7, we show the results for the two direct CP asymmetries
in (\ref{ACPs}) and (\ref{ACP2}), both for the realistic and for 
the conservative parameter sets. These results confirm the general 
observations made above. For the realistic parameter set, and with 
$\gamma$ between $47^\circ$ and $105^\circ$ as indicated by the 
global analysis of the unitarity triangle \cite{Jonnew}, we find CP 
asymmetries of up to 15\% in $B^\pm\to\pi^\pm K^0$ decays, and of 
up to 50\% in $B^\pm\to\pi^0 K^\pm$ decays. Of course, to have 
large asymmetries requires that the sines of the strong-interaction 
phases $\eta$ and $\phi$ are not small. However, this is not 
unlikely to happen. According to the left-hand plot in Figure~1,
the phase $\eta$ can take any value, and the phase $\phi$ could
quite conceivably be large due to the different decay mechanisms 
of tree- and penguin-initiated processes. We stress that there is
no strong correlation between the CP asymmetries in the two decay 
processes, because as shown in Figure~1 there is no such 
correlation between the strong-interaction phases $\eta$ and 
$\phi$. 

\subsection{Decays of neutral \boldmath$B$\unboldmath\ mesons}

Because of their dependence on the hadronic parameters 
$\varepsilon_T$, $q_C$ and $\omega_C$ entering through the sum 
$A_{1/2}+A_{3/2}$ of isospin amplitudes, the theoretical analysis 
of neutral $B\to\pi K$ decays is affected by larger hadronic 
uncertainties than that of the decays of charged $B$ mesons. 
Nevertheless, some interesting predictions regarding neutral $B$ 
decays can be made and tested experimentally. 

The expression for the direct CP asymmetry in the decays 
$B^0\to\pi^\mp K^\pm$ is
\begin{equation}
   A_{\rm CP}(\pi^- K^+) = \frac{2\sin\gamma}{R}\,
   \frac{\varepsilon_T(\sin\tilde\phi-\varepsilon_T\,q_C\sin\omega_C)
         + \varepsilon_a\,[\sin\eta-\varepsilon_T\,q_C
                           \sin(\tilde\phi-\eta+\omega_C)]}
        {1-2\varepsilon_a\cos\eta\cos\gamma+\varepsilon_a^2} \,,
\end{equation}
where $\tilde\phi=\phi_T-\phi_P$. This result reduces to (\ref{ACP2}) 
under the replacements $q_C\to\delta_{\rm EW}$, $\omega_C\to 0$, 
$R\to R_*^{-1}$, and $\varepsilon_T\to\varepsilon_{3/2}$. The 
corresponding expression for the direct CP asymmetry in the decays 
$B^0\to\pi^0 K^0$ and $\bar B^0\to\pi^0\bar K^0$ is more complicated 
and will not be presented here. Below, we shall derive an exact 
relation between the various asymmetries, which can be used to 
compute $A_{\rm CP}(\pi^0 K^0)$.

Gronau and Rosner have emphasized that one expects 
$A_{\rm CP}(\pi^- K^+)\approx A_{\rm CP}(\pi^0 K^+)$, and that 
one could thus combine the data samples for these decays to enhance 
the statistical significance of an early signal of direct CP 
violation \cite{GR98r}. We can easily understand the argument behind 
this observation using our results. Neglecting the small rescattering 
contributions proportional to $\varepsilon_a$ for simplicity, we find
\begin{equation}
   \frac{ A_{\rm CP}(\pi^- K^+)}{ A_{\rm CP}(\pi^0 K^+)}
   \simeq \frac{1}{R_* R}\,
   \frac{\varepsilon_T(\sin\tilde\phi-\varepsilon_T\,q_C\sin\omega_C)}
        {\varepsilon_{3/2}\sin\phi} 
   \simeq \frac{1}{R_* R}\,\frac{\varepsilon_T}{\varepsilon_{3/2}} \,.
\end{equation}
In the last step, we have used that the electroweak penguin
contribution is very small because it is suppressed by an additional
factor of $\varepsilon_T$, and that the strong-interaction phases 
$\phi$ and $\tilde\phi$ are strongly correlated, as follows from the
right-hand plot in Figure~1. Numerically, the right-hand side turns 
out to be close to 1 for most of parameter space. This is evident 
from the left-hand plot in Figure~8, which confirms that there is 
indeed a very strong correlation between the CP asymmetries in the 
decays $B^0\to\pi^\mp K^\pm$ and $B^\pm\to\pi^0 K^\pm$, in agreement 
with the argument given in \cite{GR98r}. Combining the data samples 
for these decays collected by the CLEO experiment, one may have a 
chance for observing a statistically significant signal for the 
first direct CP asymmetry in $B$ decays before the operation of the 
asymmetric $B$ factories.

\FIGURE{\epsfig{file=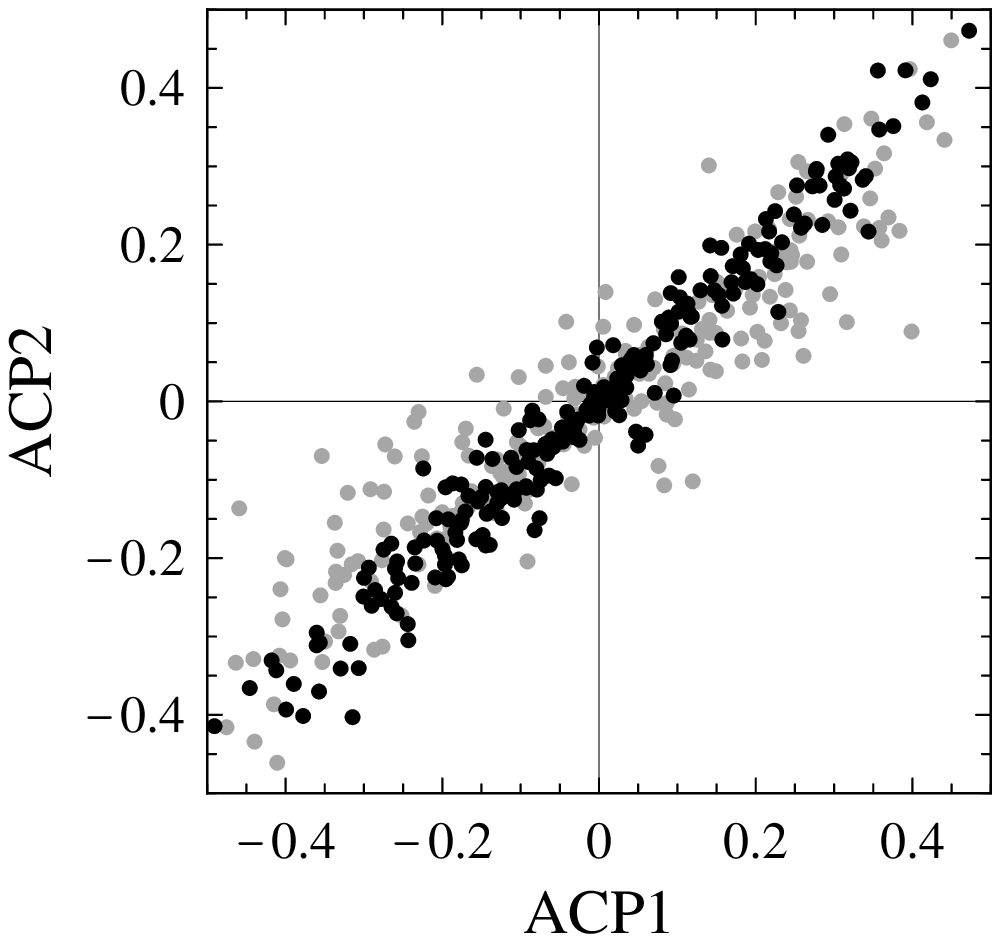,width=7.2cm}
\epsfig{file=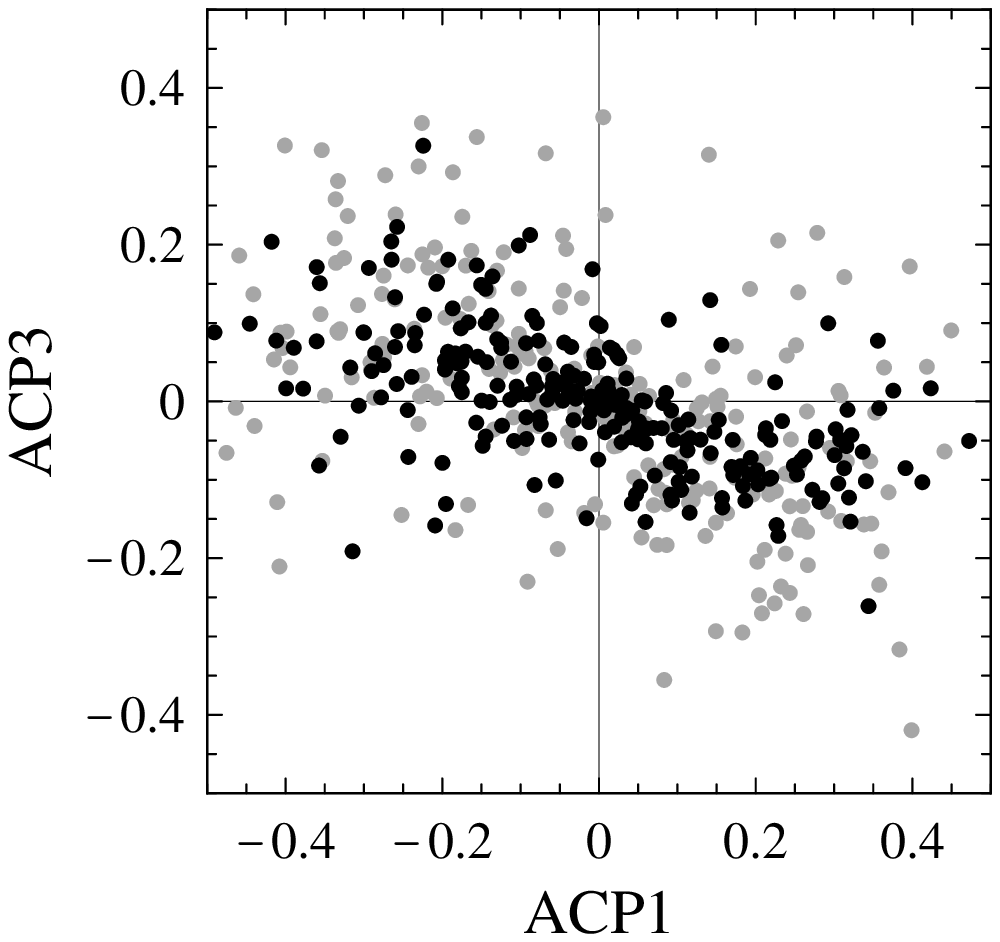,width=7.2cm} 
\caption{Correlation between the direct CP asymmetries 
$\mbox{ACP1}=A_{\rm CP}(\pi^0 K^+)$, 
$\mbox{ACP2}=A_{\rm CP}(\pi^- K^+)$, and 
$\mbox{ACP3}=A_{\rm CP}(\pi^0 K^0)$.}}

The decays $B^0\to\pi^0 K^0$ and $\bar B^0\to\pi^0\bar K^0$ have
not yet been observed experimentally, but the CLEO Collaboration
has presented an upper bound on their CP-averaged branching ratio
of $4.1\times 10^{-5}$ \cite{CLEO}. In analogy with (\ref{Rdef}),
we define the ratios
\begin{eqnarray}
   R_0 &=& \frac{\tau(B^+)}{\tau(B^0)}\,  
    \frac{2[\mbox{Br}(B^0\to\pi^0 K^0)
          +\mbox{Br}(\bar B^0\to\pi^0\bar K^0)]}
         {\mbox{Br}(B^+\to\pi^+ K^0)+\mbox{Br}(B^-\to\pi^-\bar K^0)}
    \,, \nonumber\\
   R_{0*} &=&
    \frac{2[\mbox{Br}(B^0\to\pi^0 K^0)
          +\mbox{Br}(\bar B^0\to\pi^0\bar K^0)]}
         {\mbox{Br}(B^0\to\pi^- K^+)+\mbox{Br}(\bar B^0\to\pi^+ K^-)}
    = \frac{R_0}{R} \,.
\label{R0def}
\end{eqnarray}
Using our parametrizations for the different isospin amplitudes, we 
find that the ratios $R$, $R_*$ and $R_0$ obey the relations
\begin{equation}
   R_0 - R + R_*^{-1} -1 = \Delta_1 \,, \qquad
   R_0 - R\,R_* = \Delta_2 + O(\bar\varepsilon_i^3) \,,
\label{sumrules}
\end{equation}
where
\begin{eqnarray}
   \Delta_1 &=& 2\bar\varepsilon_{3/2}^2\, 
    (1-2\delta_{\rm EW}\cos\gamma+\delta_{\rm EW}^2) 
    - 2\bar\varepsilon_{3/2}\,\bar\varepsilon_T\, 
    (1-\delta_{\rm EW}\cos\gamma) \cos(\phi_T-\phi_{3/2}) \nonumber\\
   &&\mbox{}- 2\bar\varepsilon_{3/2}\,\bar\varepsilon_T\,q_C\,
    (\delta_{\rm EW}-\cos\gamma) \cos(\phi_T-\phi_{3/2}+\omega_C) \,,
    \nonumber\\
   \Delta_2 &=& \Delta_1 - 4\bar\varepsilon_{3/2}^2
    (\delta_{\rm EW}-\cos\gamma)^2\cos^2\!\phi \nonumber\\
   &&\mbox{}+ 4\bar\varepsilon_{3/2}\,\bar\varepsilon_T\,
    (\delta_{\rm EW}-\cos\gamma)\cos\phi \left[
    q_C\cos(\tilde\phi+\omega_C)-\cos\gamma\cos\tilde\phi \right] \,, 
\end{eqnarray}
and $\bar\varepsilon_T$ is defined in analogy with 
$\bar\varepsilon_{3/2}$ in (\ref{bar32}), so that 
$\bar\varepsilon_T/\varepsilon_T=\bar\varepsilon_{3/2}
/\varepsilon_{3/2}$. The first relation in (\ref{sumrules}) 
generalizes a sum rule derived by Lipkin, who neglected the terms 
of $O(\varepsilon_i^2)$ on the right-hand side as well as 
electroweak penguin contributions \cite{Lipkin}. The second 
relation is new. It follows from the fact that $R_{0*}=R_* 
+O(\varepsilon_i^2)$, which is evident since the pairs of 
decay amplitudes entering the definition of the two ratios differ
only in the isospin amplitude $A_{3/2}$. 

\FIGURE{\epsfig{file=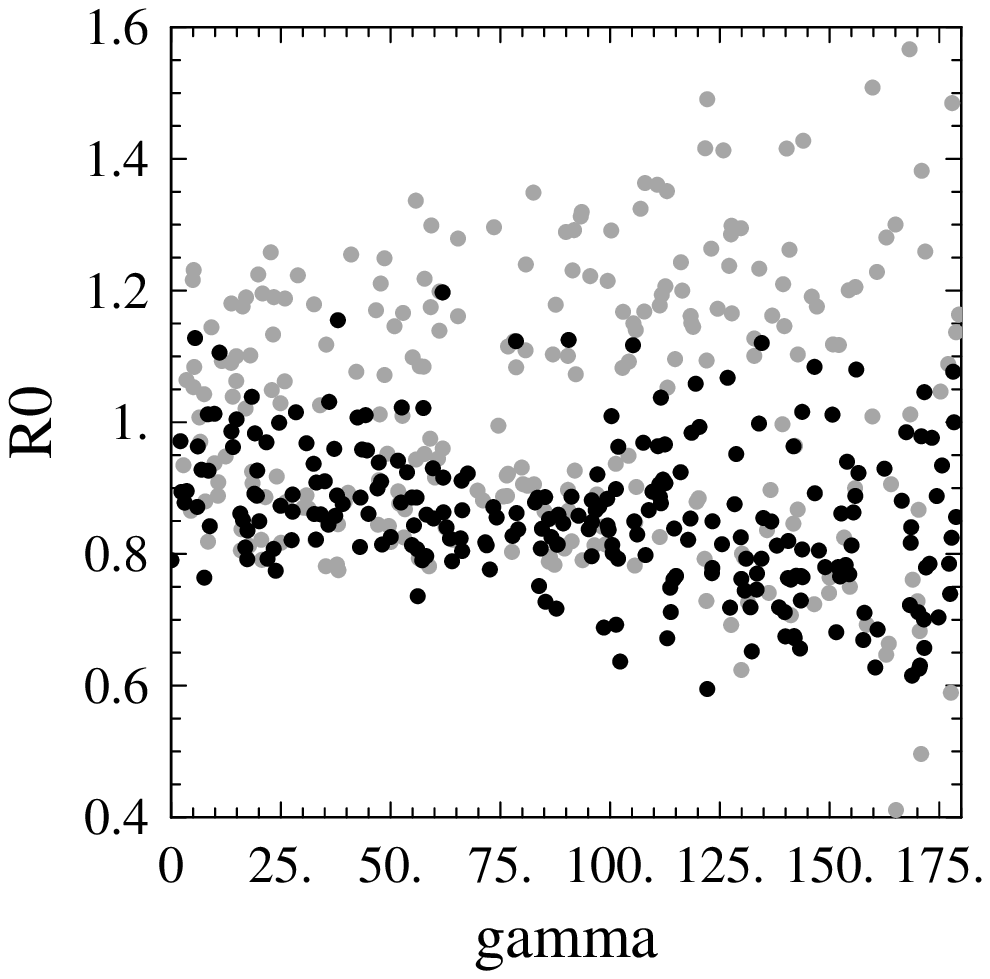,width=7.2cm}
\epsfig{file=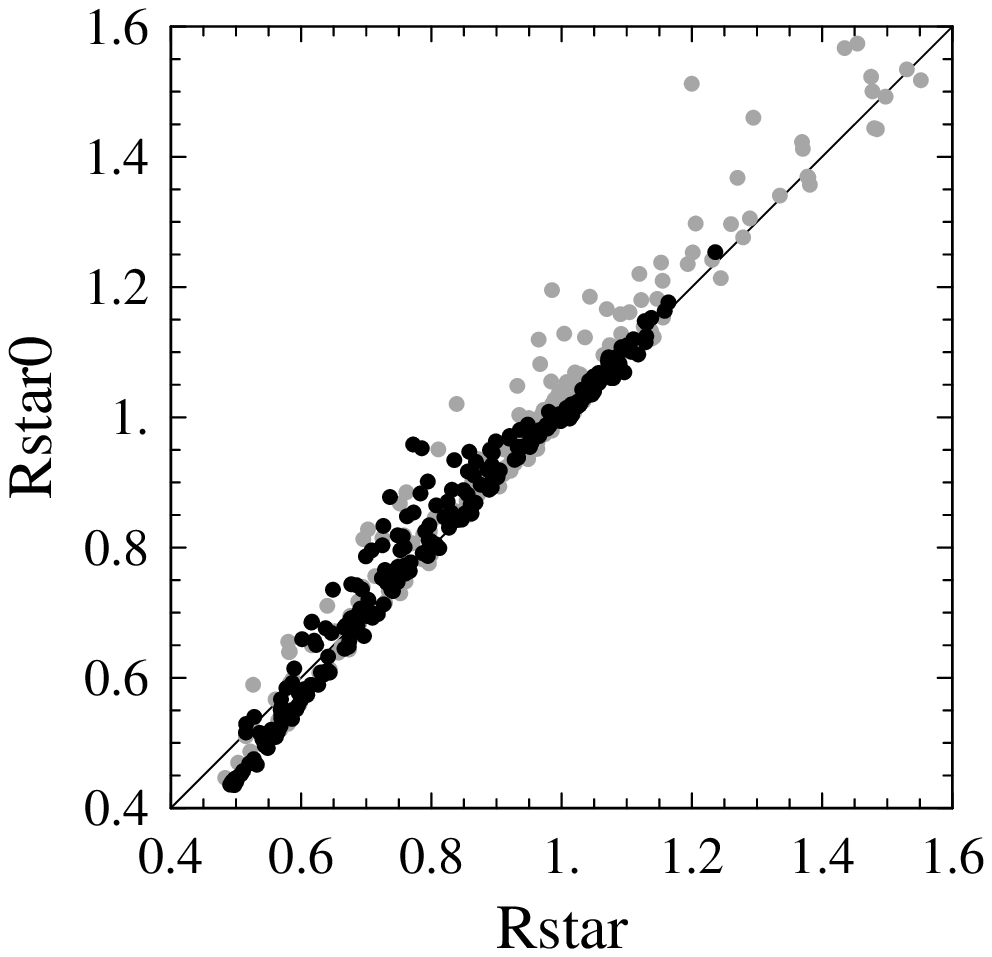,width=7.2cm} 
\caption{Ratio $R_0$ (left), and correlation between the ratios 
$R_*$ and $R_{0*}$ (right).}}

The left-hand plot in Figure~9 shows the results for the ratio $R_0$ 
versus $|\gamma|$. The dependence of this ratio on the weak phase 
turns out to be much weaker than in the case of the ratios 
$R$ and $R_*$. For the realistic parameter set we find that 
$0.7<R_0<1.0$ for most choices of strong-interaction parameters. 
Combining this with the current value of the $B^\pm\to\pi^\pm K^0$ 
branching ratio, we obtain values between 
$(0.47\pm 0.18)\times 10^{-5}$ and $(0.67\pm 0.26)\times 10^{-5}$ 
for the CP-averaged $B^0\to\pi^0 K^0$ branching ratio. The 
right-hand plot in Figure~9 shows the strong correlation between 
the ratios $R_*$ and $R_{0*}=R_0/R$, which holds with a remarkable 
accuracy over all of parameter space. 

\FIGURE{\epsfig{file=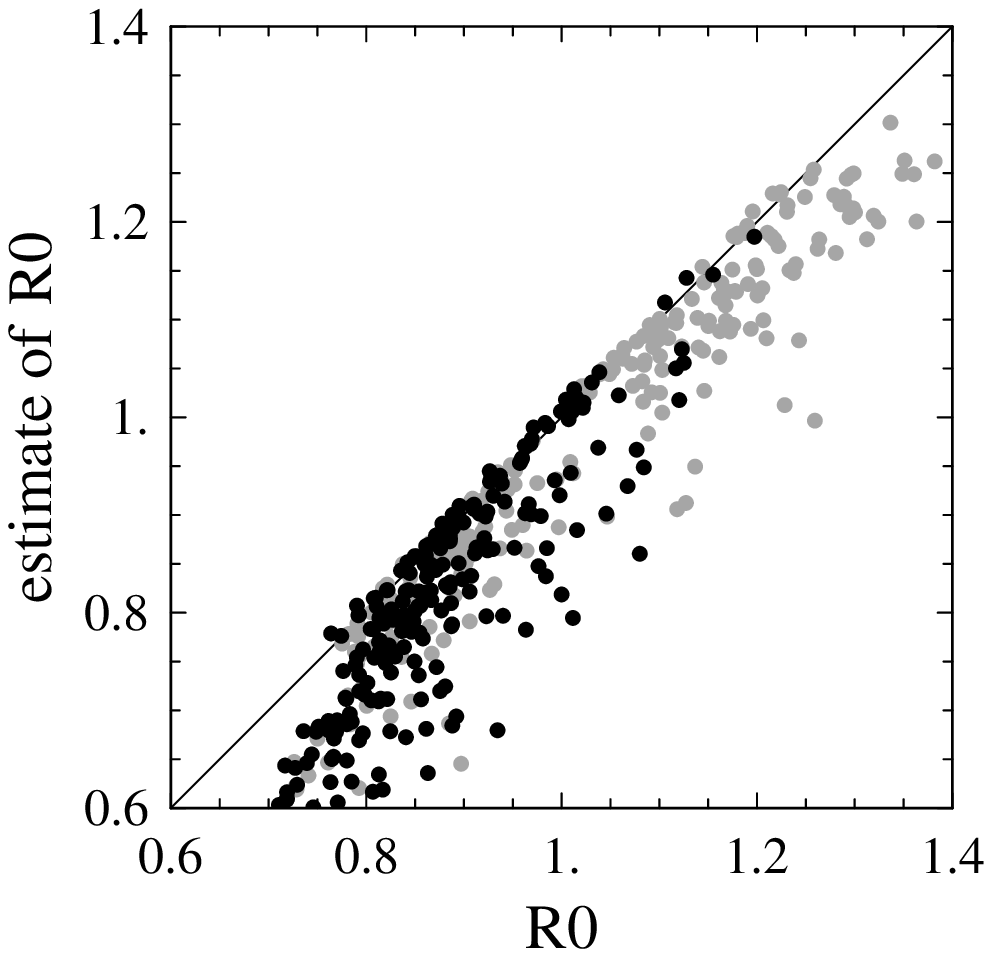,width=7.2cm}
\epsfig{file=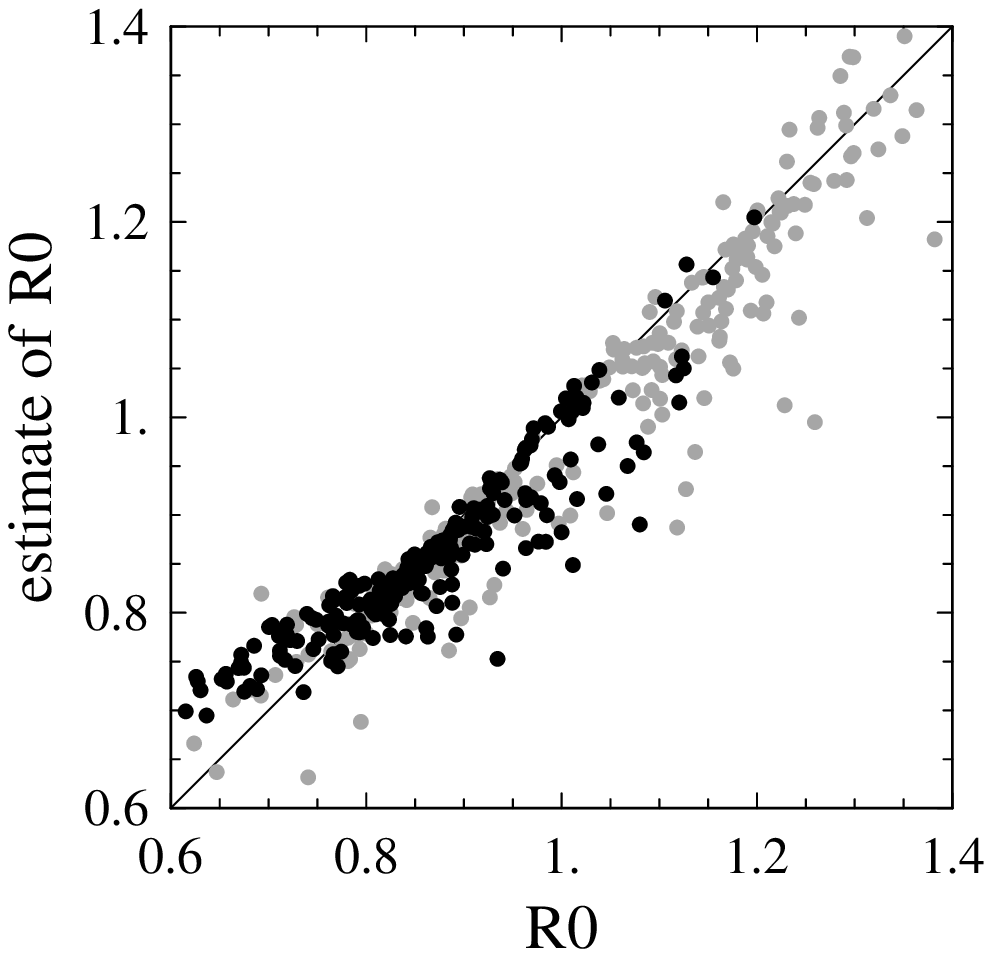,width=7.2cm} 
\caption{Esimates of $R_0$ obtained using the first (left) and
second (right) sum rule in (\protect\ref{sumrules}).}}

In Figure~10, we show the estimates of $R_0$ obtained by neglecting
the terms of $O(\bar\varepsilon_i^2)$ and higher in the two sum rules 
in (\ref{sumrules}). Using the present data for the various branching 
ratios yields to the estimates $R_0=(-0.1\pm 0.9)$ from the first 
and $R_0=(0.5\pm 0.2)$ from the second sum rule. Both results are 
consistent with the theoretical expectations for $R_0$ exhibited in 
the left-hand plot in Figure~9; however, the second estimate has a 
much smaller experimental error and, according to Figure~10, it is 
likely to have a higher theoretical accuracy. We can rewrite this 
estimate as
\begin{equation}
   \frac12 \Big[ \mbox{Br}(B^0\to\pi^0 K^0)
   + \mbox{Br}(\bar B^0\to\pi^0\bar K^0) \Big] 
   \simeq \frac{\mbox{Br}(B^\pm\to\pi^\pm K^0)\,
                \mbox{Br}(B^0\to\pi^\mp K^\pm)}
               {4\mbox{Br}(B^\pm\to\pi^0 K^\pm)} \,,
\label{pred}
\end{equation}  
where the branching ratios on the right-hand side are averaged
over CP-conjugate modes. With current data, this relation yields the
value $(0.33\pm 0.18)\times 10^{-5}$. Combining the three estimates 
for the CP-averaged $B^0\to\pi^0 K^0$ branching ratio presented 
above we arrive at the value $(0.5\pm 0.2)\times 10^{-5}$, which is 
about a factor of 3 smaller than the other three $B\to\pi K$ 
branching ratios quoted in (\ref{CLEOvals}).

We now turn to the study of the direct CP asymmetry in the decays
$B^0\to\pi^0 K^0$ and $\bar B^0\to\pi^0\bar K^0$. Using our general
parametrizations, we find the sum rule
\begin{eqnarray}
   &&A_{\rm CP}(\pi^+ K^0) - R_*^{-1}\,A_{\rm CP}(\pi^0 K^+)
    + R\,A_{\rm CP}(\pi^- K^+) - R_0\,A_{\rm CP}(\pi^0 K^0)
    \nonumber\\
   &&\quad = 2\sin\gamma\,\bar\varepsilon_{3/2}\,\bar\varepsilon_T
    \left[ \delta_{\rm EW}\sin(\phi_T-\phi_{3/2})
    - q_C\sin(\phi_T-\phi_{3/2}+\omega_C) \right] \,.
\label{ACPsum}
\end{eqnarray}
By scanning all strong-interaction parameters, we find that for the
realistic (conservative) parameter set the right-hand side takes 
values of less that 4\% (7\%) times $\sin\gamma$ in magnitude. 
Neglecting these small terms, and using the approximate equality of 
the CP asymmetries in $B^\pm\to\pi^0 K^\pm$ and 
$B^0\to\pi^\mp K^\pm$ decays as well as the second relation in 
(\ref{sumrules}), we obtain
\begin{equation}
   A_{\rm CP}(\pi^0 K^0) \simeq
   - \frac{1 - R\,R_*}{R\,R_*^2}\,A_{\rm CP}(\pi^0 K^+)
   + \frac{A_{\rm CP}(\pi^+ K^0)}{R\,R_*} \,.
\end{equation}
The first term is negative for most choices of parameters and 
would dominate if the CP aymmetry in $B^\pm\to\pi^0 K^\pm$ decays
would turn out to be large. We therefore expect a weak 
anticorrelation between $A_{\rm CP}(\pi^0 K^0)$ and 
$A_{\rm CP}(\pi^+ K^0)$, which is indeed exhibited in the right-hand 
plot in Figure~9. 

For completeness, we note that in the decays $B^0$, 
$\bar B^0\to\pi^0 K_S$ one can also study mixing-induced CP 
violation, as has been emphasized recently in \cite{BFnew}. Because 
of the large hadronic uncertainties inherent in the calculation of 
this effect, we do not study this possibility further.

\section{Determination of \boldmath$\gamma$ from $B^\pm\to\pi K$, 
$\pi\pi$ decays\unboldmath}
\label{sec:6}

Ultimately, one would like not only to derive bounds on the weak
phase $\gamma$, but to measure this parameter from a study of
CP violation in $B\to\pi K$ decays. However, as we have pointed 
out in Section~\ref{sec:2}, this is not a trivial undertaking
because even perfect measurements of all eight $B\to\pi K$
branching ratios would not suffice to eliminate all hadronic
parameters entering the parametrization of the decay amplitudes.

Because of their theoretical cleanness, the decays of charged
$B$ mesons are best suited for a measurement of $\gamma$. In
\cite{us2}, we have described a strategy for achieving this 
goal, which relies on the measurements of the CP-averaged
branching ratios for the decays $B^\pm\to\pi^\pm K^0$ and 
$B^\pm\to\pi^\pm\pi^0$, as well as of the individual branching 
ratios for the decays $B^+\to\pi^0 K^+$ and $B^-\to\pi^0 K^-$,
i.e., the direct CP asymmetry in this channel. This method is
a generalization of the Gronau--Rosner--London (GRL) approach 
for extracting $\gamma$ \cite{GRL}. It includes the contributions 
of electroweak penguin operators, which had previously been 
argued to spoil the GRL method \cite{DeHe,GHLR2}. 

The strategy proposed in \cite{us2} relies on the dynamical
assumption that there is no CP-violating contribution to the 
$B^\pm\to\pi^\pm K^0$ decay amplitudes, which is equivalent to 
saying that the rescattering effects parametrized by the quantity
$\varepsilon_a$ in (\ref{ampl1}) are negligibly small. It is
evident from the left-hand plot in Figure~1 that this assumption 
is indeed justified in a large region of parameter space. Here, 
we will refine the approach and investigate the theoretical 
uncertainty resulting from $\varepsilon_a\ne 0$. As a side 
product, we will show how nontrivial information on the 
strong-interaction phase difference $\phi=\phi_{3/2}-\phi_P$ can 
be obtained along with information on $\gamma$.

To this end, we consider in addition to the ratio $R_*$ the 
CP-violating observable
\begin{equation}
   \widetilde A \equiv \frac{A_{\rm CP}(\pi^0 K^+)}{R_*}
   - A_{\rm CP}(\pi^+ K^0) = 2\sin\gamma\,\bar\varepsilon_{3/2}\,
   \frac{\sin\phi-\varepsilon_a\,\delta_{\rm EW}\sin(\phi-\eta)}
        {\sqrt{1-2\varepsilon_a\cos\eta\cos\gamma+\varepsilon_a^2}}
   \,.
\label{Atil}
\end{equation}
The purpose of subtracting the CP asymmetry in the decays 
$B^\pm\to\pi^\pm K^0$ is to eliminate the contribution of 
$O(\varepsilon_a)$ in the expression for $A_{\rm CP}(\pi^0 K^+)$ 
given in (\ref{ACP2}). A measurement of this asymmetry is the new 
ingredient in our approach with respect to that in \cite{us2}. With 
the definition of $\widetilde A$ as given above, the rescattering 
effects parametrized by $\varepsilon_a$ are suppressed by an 
additional factor of $\bar\varepsilon_{3/2}$ and are thus expected 
to be very small. As shown in Section~\ref{sec:4}, the same is true 
for the ratio $R_*$. Explicitly, we have
\begin{eqnarray}
   R_*^{-1} &=& 1 + 2\bar\varepsilon_{3/2}\,\cos\phi\,
    (\delta_{\rm EW}-\cos\gamma) + \bar\varepsilon_{3/2}^2\,
    (1-2\delta_{\rm EW}\cos\gamma+\delta_{\rm EW}^2)
    + O(\bar\varepsilon_{3/2}\,\varepsilon_a) \,, \nonumber\\
   \widetilde A &=& 2\sin\gamma\,\bar\varepsilon_{3/2}\,\sin\phi
    + O(\bar\varepsilon_{3/2}\,\varepsilon_a) \,.
\end{eqnarray}

These equations define contours in the $(\gamma,\phi)$ plane. When 
higher-order terms are kept, these contours become 
narrow bands, the precise shape of which depends on the values of 
the parameters $\bar\varepsilon_{3/2}$ and $\delta_{\rm EW}$. 
In the limit $\varepsilon_a=0$ the procedure described
here is mathematically equivalent to the construction proposed in 
\cite{us2}. There, the errors on $\cos\gamma$ resulting from
the variation of the input parameters have been discussed in 
detail. For a typical example, where $\gamma=76^\circ$ and 
$\phi=20^\circ$, we found that the uncertainties resulting from a 
15\% variation of $\bar\varepsilon_{3/2}$ and
$\delta_{\rm EW}$ are $\cos\gamma=0.24\pm 0.09\pm 0.09$, 
correspondig to errors of $\pm 5^\circ$ each on the extracted
value of $\gamma$.  

Our focus here is to evaluate the additional uncertainty resulting
from the rescattering effects parametrized by $\varepsilon_a$ and 
$\eta$. For given values of $\bar\varepsilon_{3/2}$, 
$\delta_{\rm EW}$, $\varepsilon_a$, $\eta$, and $\gamma$, the exact 
results for $R_*$ in (\ref{R*}) and $\widetilde A$ in (\ref{Atil}) 
can be brought into the generic form $A\cos\phi+B\sin\phi=C$, where
in the case of $R_*$
\begin{eqnarray}
   A &=& 2\bar\varepsilon_{3/2}\,
    \frac{\delta_{\rm EW}-\cos\gamma+\varepsilon_a\cos\eta\,
          (1-\delta_{\rm EW}\cos\gamma)}
         {\sqrt{1-2\varepsilon_a\cos\eta\cos\gamma+\varepsilon_a^2}}
    \,, \nonumber\\
   B &=& 2\bar\varepsilon_{3/2}\,
    \frac{\varepsilon_a\sin\eta\,(1-\delta_{\rm EW}\cos\gamma)}
         {\sqrt{1-2\varepsilon_a\cos\eta\cos\gamma+\varepsilon_a^2}}
    \,, \nonumber\\
   C &=& R_*^{-1} - 1 - \bar\varepsilon_{3/2}^2\,
    (1-2\delta_{\rm EW}\cos\gamma+\delta_{\rm EW}^2) \,,
\end{eqnarray}
whereas for $\widetilde A$
\begin{eqnarray}
   A &=& 2\bar\varepsilon_{3/2}\,
    \frac{\varepsilon_a\,\delta_{\rm EW}\sin\eta}
         {\sqrt{1-2\varepsilon_a\cos\eta\cos\gamma+\varepsilon_a^2}}
    \,, \nonumber\\
   B &=& 2\bar\varepsilon_{3/2}\,
    \frac{1-\varepsilon_a\,\delta_{\rm EW}\cos\eta}
         {\sqrt{1-2\varepsilon_a\cos\eta\cos\gamma+\varepsilon_a^2}}
    \,, \nonumber\\
   C &=& \frac{\widetilde A}{\sin\gamma} \,.
\end{eqnarray}
The two solutions for $\cos\phi$ are given by
\begin{equation}
   \cos\phi = \frac{A C\pm B\sqrt{A^2+B^2-C^2}}{A^2+B^2} \,.
\end{equation}
The physical solutions must be such that $\cos\phi$ is real and its 
magnitude less than 1.

\EPSFIGURE{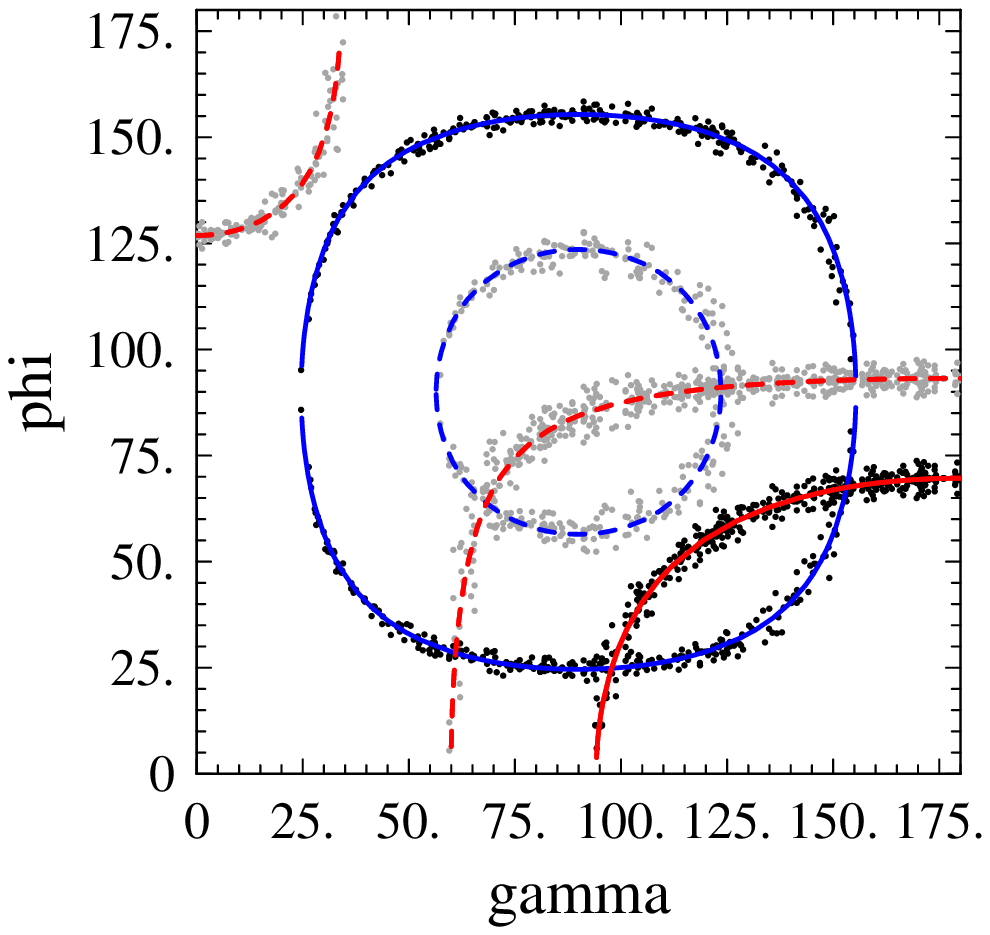,width=7.2cm}
{Contour plots for the quantities $R_*$ (red ``hyperbolas'') and 
$\widetilde A$ (blue ``circles''). The scatter plots show the 
results including rescattering effects, while the lines refer to 
$\varepsilon_a=0$. The solid curves correspond to the contours for 
$R_*=0.7$ and $\widetilde A=0.2$, the dashed ones to $R_*=0.9$ and 
$\widetilde A=0.4$.}

In Figure~12, we show the resulting contour bands obtained 
by keeping $\bar\varepsilon_{3/2}=0.24$ and $\delta_{\rm EW}=0.64$ 
fixed to their central values, while the rescattering 
parameters are scanned over the ranges $0<\varepsilon_a<0.08$ 
and $-180^\circ<\eta<180^\circ$. Assuming that $\sin\gamma>0$ as 
suggested by the global analysis of the unitarity triangle, the 
sign of $\widetilde A$ determines the sign of $\sin\phi$. In the 
plot, we assume without loss of generality that $0^\circ\le\phi\le 
180^\circ$. For instance, if $R_*=0.7$ and $\widetilde A=0.2$, 
then the two solutions are $(\gamma,\phi)\approx
(98^\circ,25^\circ)$ and $(\gamma,\phi)\approx
(153^\circ,67^\circ)$, only the first of which is allowed 
by the upper bound $\gamma<105^\circ$ following from the 
global analysis of the unitarity triangle \cite{Jonnew}.
It is evident that the contours are rather insensitive to the 
rescattering effects parametrized by $\varepsilon_a$ and $\eta$. 
The error on $\gamma$ due to these effects is about $\pm 5^\circ$, 
which is similar to the errors resulting from the theoretical 
uncertainties in the parameters $\bar\varepsilon_{3/2}$ and 
$\delta_{\rm EW}$. The combined theoretical uncertainty is 
of order $\pm 10^\circ$ on the extracted value of $\gamma$.  

To summarize, the strategy for determining $\gamma$ would be as 
follows: From measurements of the CP-averaged branching ratio for 
the decays $B^\pm\to\pi^\pm\pi^0$, $B^\pm\to\pi^\pm K^0$ and 
$B^\pm\to\pi^0 K^\pm$, the ratio $R_*$ and the parameter 
$\bar\varepsilon_{3/2}$ are determined using (\ref{Rdef}) and 
(\ref{epsexp}), respectively. Next, from measurements of the rate 
asymmetries in the decays $B^\pm\to\pi^\pm K^0$ and 
$B^\pm\to\pi^0 K^\pm$ the quantity $\widetilde A$ is determined. 
From the contour plots for the quantities $R_*$ and $\widetilde A$ 
the phases $\gamma$ 
and $\phi$ can then be extracted up to discrete ambiguities. In 
this determination one must account for theoretical uncertainties 
in the values of the parameters $\bar\varepsilon_{3/2}$ and 
$\delta_{\rm EW}$, as well as for rescattering effects 
parametrized by $\varepsilon_a$ and $\eta$. Quantitative estimates 
for these uncertainties have been given above.

\section{Conclusions}
\label{sec:7}

We have presented a model-independent, glo\-bal analysis of the
rates and direct CP asymmetries for the rare two-body decays
$B\to\pi K$. The theoretical description exploits the flavour
symmetries of the strong interactions and the structure of the 
low-energy effective weak Hamiltonian. Isospin symmetry is used
to introduce a minimal set of three isospin amplitudes. The
explicit form of the effective weak Hamiltonian in the
Standard Model is used to simplify the isovector part of the 
interaction. Both the numerical smallness of certain Wilson 
coefficient functions and the Dirac and colour structure of 
the local operators are relevant in this context. Finally, the 
$U$-spin subgroup of flavour SU(3) symmetry is used to simplify
the structure of the isospin amplitude $A_{3/2}$ referring to
the decay $B\to(\pi K)_{I=3/2}$. In the limit of exact $U$-spin 
symmetry, two of the four parameters describing this amplitude
(the relative magnitude and strong-interaction phase of 
electroweak penguin and tree contributions) can be calculated 
theoretically, and one additional parameter (the overall strength 
of the amplitude) can be determined experimentally from a 
measurement of the CP-averaged branching ratio for 
$B^\pm\to\pi^\pm\pi^0$ decays. What remains is a single unknown 
strong-interaction phase. The SU(3)-breaking corrections to these 
results can be calculated in the generalized factorization 
approximation, so that theoretical limitations enter only at the
level of nonfactorizable SU(3)-breaking effects. However, since
we make use of SU(3) symmetry only to derive relations for
amplitudes referring to isospin eigenstates, we do not expect
gross failures of the generalized factorization hypothesis. 
We stress that the theoretical simplifications used in our
analysis are the only ones rooted on first principles of QCD. 
Any further simplification would have to rest on model-dependent
dynamical assumptions, such as the smallness of certain flavour
topologies with respect to others.

We have introduced a general parametrization of the decay 
amplitudes, which makes maximal use of these theoretical 
constraints but is otherwise completely general. In particular, 
no assumption is made about strong-interaction phases. With the 
help of this parametrization, we have performed a global analysis 
of the branching ratios and direct CP asymmetries in the various 
$B\to\pi K$ decay modes, with particular emphasis on the impact 
of hadronic uncertainties on methods to learn about the weak 
phase $\gamma=\mbox{arg}(V_{ub}^*)$ of the unitarity triangle.
The main phenomenological implications of our results can be 
summarized as follows:

\begin{itemize}
\item
There can be substantial corrections to the Fleischer--Mannel bound 
on $\gamma$ from enhanced electroweak penguin contributions, which
can arise in the case of a large strong-interaction phase difference
between $I=\frac 12$ and $I=\frac 32$ isospin amplitudes. Whereas 
these corrections stay small (but not negligible) if one restricts 
this phase difference to be less than $45^\circ$, there can be 
large violations of the bound if the phase difference is allowed to
be as large as $90^\circ$.
\item
On the contrary, rescattering effects play a very minor role in the
bound on $\gamma$ derived from a measurement of the ratio $R_*$ of
CP-averaged $B^\pm\to\pi K$ branching ratios. They can be included
exactly in the bound and enter through a parameter $\varepsilon_a$, 
whose value is less than 0.1 even under very conservative conditions.
Including these effects weakens the bounds on $\gamma$ by less than
$5^\circ$. We have generalized the result of our previous work 
\cite{us}, where we derived a bound on $\cos\gamma$ to linear order 
in an expansion in the small quantity $\bar\varepsilon_{3/2}$. Here 
we refrain from making such an approximation; however, we confirm our 
previous claim that to make such an expansion is justified (i.e., 
it yields a conservative bound) provided that the current 
experimental value of $R_*$ does not change by more than one 
standard deviation. The main result of our analysis is given in 
(\ref{exact}), which shows the exact result for the maximum value 
of the ratio $R_*$ as a function of the parameters 
$\delta_{\rm EW}$, $\bar\varepsilon_{3/2}$, and $\varepsilon_a$. 
The first parameter describes electroweak penguin contributions and 
can be calculated theoretically. The second parameter can be 
determined experimentally from the CP-averaged branching ratios for 
the decays $B^\pm\to\pi^\pm\pi^0$ and $B^\pm\to\pi^\pm K^0$. We 
stress that the definition of $\bar\varepsilon_{3/2}$ is such that 
it includes exactly possible rescattering contributions to the 
$B^\pm\to\pi^\pm K^0$ decay amplitudes. The third parameter 
describes a certain class of rescattering effects and can be 
constrained experimentally once the CP-averaged 
$B^\pm\to K^\pm\bar K^0$ branching ratio has been measured. However, 
we have shown that under rather conservative assumptions 
$\varepsilon_a<0.1$.
\item
The calculable dependence of the 
$B^\pm\to\pi K$ decay amplitudes on the electroweak penguin 
contribution $\delta_{\rm EW}$ offers a window to New Physics. In
many generic extensions of the Standard Model such as multi-Higgs
models, we expect deviations from the value 
$\delta_{\rm EW}=0.64\pm 0.15$ predicted by the Standard Model. 
We have derived a lower bound on $\delta_{\rm EW}$ as a function
of the value of the ratio $R_*$ and the maximum value for $\gamma$
allowed by the global analysis of the unitarity triangle. If it
would turn out that this value exceeds the Standard Model 
prediction by a significant amount, this would be strong evidence
for New Physics. In particular, we note that if the current
central value $R_*=0.47$ would be confirmed, the value of
$\delta_{\rm EW}$ would have to be at least twice its standard
value.
\item
We have studied in detail the potential of the various $B\to\pi K$ 
decay modes for showing large direct CP violation and investigated 
the correlations between the various asymmetries. Although in 
general the theoretical predictions suffer from the fact that an 
overall strong-interaction phase difference is unknown, we 
conclude that there is a fair chance for observing large direct
CP asymmetries in at least some of the decay channels. More 
specifically, we find that the direct CP asymmetries in the decays
$B^\pm\to\pi^0 K^\pm$ and $B^0\to\pi^\mp K^\pm$ are almost fully
correlated and can be up to 50\% in magnitude for realistic 
parameter choices. The direct CP asymmetry in the decays 
$B^0\to\pi^0 K^0$ and $\bar B^0\to\pi^0\bar K^0$ tends to be 
smaller by about a factor of 2 and anticorrelated in sign. 
Finally, the asymmetry in the decays $B^\pm\to\pi^\pm K^0$ is
smaller and uncorrelated with the other asymmetries. For realistic
parameter choices, we expect values of up to 15\% for this 
asymmetry.
\item
We have derived sum rules for the branching ratio and direct CP 
asymmetry in the decays $B^0\to\pi^0 K^0$ and 
$\bar B^0\to\pi^0\bar K^0$. A rather clean prediction for the
CP-averaged branching ratio for these decays in given in 
(\ref{pred}). We expect a value of $(0.5\pm 0.2)\times 10^{-5}$
for this branching ratio, which is about a factor of 3 less than 
the other $B\to\pi K$ branching ratios.
\item
Finally, we have presented a method for determining the weak phase 
$\gamma$ along with the strong-interaction phase difference $\phi$ 
from measurements of $B^\pm\to\pi K$, $\pi\pi$ branching ratios, 
all of which are of order $10^{-5}$. This method generalizes an 
approach proposed in \cite{us2} to include rescattering corrections 
to the $B^\pm\to\pi^\pm K^0$ decay amplitudes. We find that the 
uncertainty due to rescattering effects is about $\pm 5^\circ$ on 
the extracted value of $\gamma$, which is similar to the errors 
resulting from the theoretical uncertainties in the parameters 
$\bar\varepsilon_{3/2}$ and $\delta_{\rm EW}$. The combined 
theoretical uncertainty in our method is of order $\pm 10^\circ$.  
\end{itemize}

A global analysis of branching ratios and direct CP asymmetries in 
rare two-body decays of $B$ mesons can yield interesting information 
about fundamental parameters of the flavour sector of the Standard 
Model, and at the same time provides a window to New Physics. Such 
an analysis should therefore be a central focus of the physics 
program of the $B$ factories, which in many respects is complementary 
to the time-dependent studies of CP violation in neutral $B$ decays 
into CP eigenstates.

\acknowledgments
This is my last paper as a member of the CERN Theory Division. 
It is a pleasure to thank my colleagues for enjoyful 
interactions during the past five years. I am very grateful to
Guido Altarelli, Martin Beneke, Gian Giudice, Michelangelo Mangano, 
Paolo Nason and, especially, to Alex Kagan for their help in a 
difficult period. I also wish to thank Andrzej Buras, Guido 
Martinelli, Chris Sachrajda, Berthold Stech, Jack Steinberger and 
Daniel Wyler for their support. It is a special pleasure to thank
Elena, Jeanne, Marie-Noelle, Michelle, Nannie and Suzy for 
thousands of smiles, their friendliness, patience and help. 
Finally, I wish to the CERN Theory Division that its structure may 
change in such a way that one day it can be called a Theory 
{\em Group}.

\end{document}